\documentclass[a4paper,11pt,x11names]{article}
\pdfoutput=1
\usepackage{jheppub}
\usepackage{graphicx}
\usepackage{amsmath}
\usepackage{amsfonts}
\usepackage{mathrsfs}
\usepackage{amssymb}
\usepackage{slashed}
\usepackage[colorlinks=true,linkcolor=blue]{hyperref}
\usepackage{bm}
\usepackage{empheq}
\usepackage[super]{nth}
\usepackage{color}
\def\sss{\scriptscriptstyle}

\def\Eq#1{Eq.~\eqref{#1}}
\def\Re{\,\mathrm{Re}\,}
\def\Im{\,\mathrm{Im}\,}
\def\nn{\nonumber}

\def\CC{\mathcal{C}}
\def\OO{\mathcal{O}}
\def\nf{N_{\mathrm{f}}}
\def\ffhc{\mathrm{f\bar{f}hc}}

\def\Qmax{\mathcal{Q}_{\mathrm{max}}}

\def\lra{\leftrightarrow}

\def\PiL{\Pi_{00}}
\def\PiT{\Pi_{\mathrm T}}
\def\Trace{\mathrm{Tr}\,}
\def\onetwo{{1\lra2}}
\def\twotwo{{2\lra2}}

\def\Qmax{Q_{\mathrm{max}}}

\def\muEQCD{\overline\mu_{_{\mathrm{EQCD}}}}

\def\p{{\bm p}}
\def\q{{\bm q}}
\def\k{{\bm k}}

\def\bh{{\textbf{h}}}
\def\bb{\textbf{b}}
\def\qp{q_\perp}

\def\cf{C_{\mathrm F}}

\newcommand{\Tint}[1]{{\hbox{$\sum$}\!\!\!\!\!\!\!\int\,}_{\!\!\!\!\raise-0.9ex\hbox{$\scriptstyle{#1}$}}}

\def \bfnabla {\boldsymbol{\nabla}}

\def \cc {\mathcal{C}}
\def \als {\alpha_{\mathrm{s}}}

\def\md {m_{\sss D}}
\def\mD{\md}
\def\semi {\mathrm{semi}}
\def\st{\begin{equation}}
\def\stp{\end{equation}}
\def\A{{\mathcal C}}

\title{Hot and Dense QCD Shear Viscosity at (almost) NLO}
\author{Isabella Danhoni, Guy D. Moore}
\affiliation{Institut f\"ur Kernphysik, Technische Universit\"at Darmstadt\\
Schlossgartenstra{\ss}e 2, D-64289 Darmstadt, Germany}
\emailAdd{idanhoni@theorie.ikp.physik.tu-darmstadt.de,guy.moore@physik.tu-darmstadt.de}

\abstract{The next-to-leading order weak-coupling shear viscosity of QCD was computed 6 years ago.
 However, these results have never been applied at finite baryon chemical potential $\mu$, even though intermediate-energy heavy ion collisions and merging neutron stars may explore the Quark-Gluon Plasma in a regime where baryon chemical potentials are large.
 Here, we extend the next-to-leading order shear viscosity calculations to finite $\mu$, and we show that, while the convergence of the weak-coupling expansion is questionable for achievable plasmas, it is somewhat better at $\mu > T$ than at $\mu=0$.

}


\date{\today}

\begin{document}

\maketitle

\section{Introduction}
\label{sec:intro}

The deconfined phase of hot and dense QCD is known as the quark-gluon plasma (QGP). Evidence shows that this phase was present in the early universe and that it has been recreated at microscopic scales in heavy ion collision experiments\cite{Grefa:2022sav}.
These results have sparked large interest in this subject over the past few decades, and experiments indicate that the QGP in such collisions is a strongly coupled fluid\cite{PHENIX:2004vcz,STAR:2005gfr,ALICE:2010suc} with behavior close to that of a perfect fluid\cite{Heinz:2013th,Luzum:2013yya,JETSCAPE:2020mzn}.
On the other hand, sufficiently early in the history of the early Universe, the running coupling should have been small enough that a perturbative treatment may be valid.
For thermodynamical quantities, suitably resummed perturbative calculations work starting at temperatures a few times that of the chiral symmetry restoration transition\cite{Laine:2003ay,Hietanen:2008tv}.


In lower-energy collisions, quark-gluon plasma is generated with a significant net baryon number, such that the baryon chemical potential $\mu_B$ is several times higher than the temperature.
This regime cannot be directly studied on the lattice due to the sign problem (or complex action problem).
For a chemical potential $\mu_B\neq 0$ the fermion determinant (or Monte Carlo action) becomes complex, which ruins the probabilistic interpretation of this quantity used in lattice QCD calculations\cite{Goy:2016egl}.
For this reason, the high-density region of the QCD phase diagram is much less well understood than the $\mu_B=0$ axis.
As experimental efforts focus on bringing clarity to this region, such as RHIC-BES \cite{STAR:2014egu}, HADES\cite{HADES:2019auv} and CBM at
FAIR\cite{Friese:2006dj}, the need for parallel theoretical development becomes clear.

At high temperatures, we claim that the perturbative series should work better at a high chemical potential, such that $\mu_B > T > T_c$, than in the case of $\mu_B=0$\cite{Danhoni:2022xmt}. This is a consequence of the fact that scattering from quarks is enhanced by a factor of $\mu_B^2 / T^2$.
The failure of perturbation theory in hot QCD occurs before the coupling becomes large, due to the strongly-coupled nature of soft gluons.
These play a central role in the breakdown of the perturbative expansion for the shear viscosity \cite{Ghiglieri:2018dib}.
But at high chemical potential, where quarks play a larger role, one expects the perturbative series to be better behaved.
This means that perturbation theory has a better chance of working at high $\mu/T$ than at $\mu=0$.
Although the $(\mu,T)$ values which can be explored in experiment are unlikely to extend into the region where perturbation theory is directly predictive, the perturbative results can still be used in modeling the $(\mu,T)$ behavior of the shear viscosity phenomenologically, as was recently done using leading-log results in Ref.~\cite{Danhoni:2024kgi}.
Essentially, perturbation theory serves as a limiting behavior so that $\eta/s$ can be interpolated based on high-density perturbative, low-density hadron-resonance gas, and intermediate density data-driven estimates.

Previously Arnold, Moore, and Yaffe have computed the shear viscosity at leading perturbative order in the absence of a chemical potential
\cite{Arnold:2003zc}, and Ghiglieri et al have extended this calculation to ``almost'' next-to-leading order (NLO) \cite{Ghiglieri:2018dib}.
Separately, we have presented the shear viscosity at finite chemical potential at leading-log order \cite{Danhoni:2022xmt}.
The goal of this paper is to work to almost next-to-leading order at finite chemical potential, both to obtain the most reliable estimates for the viscosity in this regime that we can, and to test how perturbation theory converges in the high-$\mu$ regime.
We start in Section \ref{sec:kineticthy} by reviewing the kinetic theory and computational approach that Arnold, Moore, and Yaffe used in \cite{Arnold:2000dr,Arnold:2003zc}.
In Section \ref{sec:NLO}, we review how to extend the collision operator to next-to-leading order and describe the computational tools required. 
In Section \ref{sec:mu}, we show what changes must be made to the calculation to include chemical potentials.
The analysis of our calculations and final results are shown in
Section \ref{sec:results}.
We end in Section \ref{sec:summary} with a short discussion and summary.
But we can present the main conclusions very briefly now.
The shear viscosity, rescaled as $\eta T/(\varepsilon + P)$, is a weak function of $\mu_q/T$ (with $\mu_q=\mu_B/3$ the chemical potential of a quark).
The convergence of the perturbative series improves when
$\nf \mu_q^2 > N_c \pi^2 T^2$, but the improvement is somewhat modest.
For more complete details, see Section \ref{sec:results}.

\section{Kinetic Theory}
\label{sec:kineticthy}
In this section, we briefly summarize how shear viscosity is defined and how it can be computed to leading order (LO) in the Effective Kinetic Theory (EKT).
Shear viscosity is a property relevant for systems with space-nonuniform flow velocity, meaning that the energy and momentum densities vary through space, but in a manner which is sufficiently smooth that the system is locally near equilibrium.
In this case we can exchange the conserved densities $T^{\mu 0}$ for the local 4-velocity and energy density, defined with the Landau-Lifshitz convention \cite{Landau1987Fluid}:
$u^\mu(x,t)$ is the timelike unit eigen(four)vector of $T^{\mu\nu}$ and $\varepsilon$ is its eigenvalue,
$T^{\mu\nu} u_\nu = \varepsilon u^\mu$
with\footnote{We use a mostly-positive metric tensor $g_{\mu\nu} = \mathrm{Diag}[-1,+1,+1,+1]$,
and natural units, $c=1$, $\hbar = 1$. Capital letters $P,K$ are 4-vectors, lower case letters $\vec p,\vec k$ or $p_i,k_i$ are the space components, $p^0,k^0$ are the time components, and $p,k$ are the magnitudes of the space components.}
$u^\mu u_\mu = -1$.
For uniform $u^\mu,\varepsilon$ the system will be in equilibrium with
$T^{\mu\nu} = \varepsilon u^\mu u^\nu + P g^{\mu\nu}$,
where $P$ the pressure is determined from $\varepsilon$ by the equilibrium equation of state.

However, for nonuniform $u^\mu$ the system will depart from equilibrium and $T^{\mu\nu}$ will not follow its ideal form.
Defining the projection operator into the local rest frame,
\begin{equation}
\label{defDelta}
    \Delta^{\mu\nu} \equiv g^{\mu\nu} + u^\mu u^\mu \,,
\end{equation}
the shear tensor is defined as
\begin{equation}
\label{defsigma}
    \sigma^{\mu\nu} \equiv \nabla^\mu u^\nu + \nabla^\nu u^\mu
    - \frac{2}{3} \Delta^{\mu\nu} \nabla_\alpha u^\alpha \,.
\end{equation}
For small and slowly varying $\sigma_{ij}$, linear response theory predicts that the stress tensor should change by an amount linear in the shear tensor:  in the local rest frame,
\begin{equation}
\label{etadef}
    T_{ij} = P g_{ij} - \eta \, \sigma_{ij}
\end{equation}
with $P$ the pressure and $\eta$, the shear viscosity, defined as the coefficient of the linear response of $T_{ij}$ to nonvanishing $\sigma_{ij}$.
If the fluid flow also has a divergence $\theta = \partial_i u_i$, then an additional term $T_{ij} = \ldots - \zeta \theta \Delta_{ij}$ would be present, with $\zeta$ the bulk viscosity.
Bulk viscosity requires a different analysis \cite{Arnold:2006fz} which we will not attempt here.

Many (not all) theories have a kinetic description, and this should include nonabelian gauge theories in the weak coupling regime.
In this case one can give a coarse-grained description of the state of the system in terms of the occupancies of each field type at each point in time and phase space, $f^a(\vec p,\vec x,t)$.
In equilibrium and at lowest order in the coupling, we have
\begin{equation}
\label{f0}
    f_0^a(\vec p,\vec x,t) = \frac{1}{\exp(\gamma \beta ( p^0 - u_i p^i - \mu_a ) \mp 1 }
    = \frac{1}{\exp (\beta (-u_\mu P^\mu - \mu_a)) \mp 1}.
\end{equation}
Here $u^\mu = (\gamma , \gamma \vec v)$ is the local velocity 4-vector as before, $\gamma$ is the associated relativistic gamma-factor, and $\beta=\beta(\vec x,t) = 1/T$ is the local temperature.
The sign $\mp 1$ is $-$ for bosons and $+$ for fermions (the upper sign will always apply to bosons), and $\mu_a$ is the chemical potential for species $a$, which will be nonzero for quark species and opposite for a particle and its antiparticle.
In this work, we will consider $SU(3)$ gauge theory with $\nf$ vectorlike quark species each with the same chemical potential.

Shear viscosity, as well as other transport coefficients, is theoretically defined in a Kubo-type formulae relating them to the zero-frequency transport limit of the spectral functions of two-point correlators of the appropriate operators.
Performing a perturbative evaluation at leading order using the diagrammatic approach based directly on these Kubo formulas would require complex resummations of numerous sub-diagrams to all orders.
This has been done successfully for scalar field theory
\cite{Jeon:1994if} and for QED for electrical conductivity
\cite{Gagnon:2006hi} and shear viscosity \cite{Gagnon:2007qt},
and in each case the result proved to be the same as a kinetic description.
In the case of full QCD such a diagrammatic calculation has never been successfully attempted, but physical arguments indicate that the result should again justify a kinetic description \cite{Arnold:2002zm}.
Therefore, we adopt the linearized version of the EKT developed in \cite{Arnold:2002zm}.

One starts as usual with a Boltzmann equation
\begin{equation}
    \Big[ \vec{v}_p \cdot \frac{\partial}{\partial \vec{x}} \Big] f^a(\vec{p},\vec{x},t) = - \CC^a[f],
    \label{boltz}
\end{equation}
where $f^a$ refers to the nonequilibrium form of the statistical function,
which we linearize about the equilibrium form
\begin{equation}
\label{f1}
    f^a(\vec p,\vec x) = f_0^a(\vec p, \vec x) + f_0(1 \pm f_0)\, f_1^a(\vec p,\vec x) \,,
\end{equation}
according to the Landau-Lifshitz conventions under which $f_0$ contributes all of the energy, momentum density, and net particle numbers in the system.
The choice to normalize $f_1$ with a factor of $f_0(1\pm f_0)$ will be convenient in what follows.
For a more detailed description one can refer to \cite{Baym:1990uj} which developed the leading-log application of Boltzmann equations to the evaluation of the QCD shear viscosity,
\cite{Arnold:2003zc}, which extended the treatment to leading order and established most of our notation and techniques,
and \cite{Danhoni:2022xmt}, which extended the treatment to finite chemical potentials, albeit at leading-logarithmic order.
We will save the discussion of the collision operator $\CC^a[f]$ for the next section.

The departure from equilibrium must first arise at linear order in $\sigma_{ij}$.
Using rotational invariance, one can then write the departure from equilibrium as
\begin{equation}
\label{chidef}
    f_1^a(\vec p,\vec x) = \beta^2 X_{ij}(\vec x) \chi^a_{ij}(\vec p) = \beta^2 X_{ij}(\vec x) I_{ij}(\vec p) \chi^a(p) \,.
\end{equation}
Here $\chi^a(p)$ is a species-dependent, pure scalar function of $p$ which characterizes the departure from equilibrium in the presence of shear stress, and
\begin{align}
\label{Xdef}
    X_{ij}& \equiv \frac{1}{\sqrt{6}} \sigma_{ij}\\
    \label{IijDef}
    I_{ij}(\vec p) & \equiv \sqrt{\frac 32} \left( \hat p_i \hat p_j - \frac{1}{3} \delta_{ij} \right) \,.
\end{align}
The normalization is explained in Ref.~\cite{Arnold:2003zc} and is chosen so that $I_{ij}$ is a properly normalized projection operator.
Evaluating the space derivatives on the lefthand side of \Eq{boltz}
one finds to lowest order,
\begin{equation}
    \label{LHSBoltz}
    \vec v_p \cdot \frac{\partial}{\partial \vec x}
    f^a_0(\vec p,\vec x,t) = \beta^2 X_{ij} S^a_{ij}
\end{equation}
with
\begin{equation}
    \label{SijDef}
    S^a_{ij}(\vec p)  \equiv p^0 \, T \, f_0^a(1\pm f_0^a) I_{ij}(\vec p) \,.
\end{equation}
Using this and inserting \Eq{chidef} into the RHS, \Eq{boltz} reduces to
\begin{equation}
    \beta^2 X_{ij} \; S^a_{ij}(\vec p) = \CC^a[\beta^2 X_{ij} \, \chi^a_{ij}(\vec p)]
    \quad \mbox{and therefore} \quad
    S^a_{ij}(\vec p) = \CC \chi^a_{ij}(\vec p) ,
    \label{boltz2}
\end{equation}
where $\CC \chi^a_{ij}(\vec p)$ means the result of factoring $\beta^2 X_{ij}$ out of the collision operator.

This equation is solved using a variational method.
This strategy was used in \cite{Arnold:2000dr,Arnold:2003zc} in leading log and leading order calculations. 
First one defines an inner product on departures from equilibrium:
 \begin{equation}
 \label{innerprod}
     (f,g) = \beta^3 \sum^{\mathrm{\ffhc}}_{a} \int_{\vec{p}} f^a(\vec{p})g^a(\vec{p}) 
 \end{equation}
It can be shown that the collision operator is Hermitian with respect to this inner product.
We define
 \begin{equation}
 \label{Qdef}
     \mathcal{Q}[\chi] = \left(\chi_{ij}, S_{ij}\right)- \frac{1}{2}\left(\chi_{ij},\CC\chi_{ij}\right) \,,
 \end{equation}
which can be viewed as a functional over the space of $\chi^a(p)$ values.
That is, the functional $\mathcal{Q}$ takes its maximum value when the Boltzmann equation is satisfied.
Furthermore, its value at this maximum,
 \begin{equation}
 \label{Qmaxval}
     \Qmax=\frac{1}{2}(\chi_{ij}, S_{ij})=\frac{1}{2} (\chi_{ij},\CC\chi_{ij}) = \frac{1}{2} ( S_{ij},\CC^{-1} S_{ij} ) \,,
 \end{equation} 
determines the viscosity:
 \begin{equation}
 \label{etaQmax}
     \eta = \frac{2}{15}\Qmax.
 \end{equation}
As known from other variational problems, one can obtain highly accurate approximate results by performing a restricted extremization within a well-chosen finite dimensional subspace.
For this purpose, we expand the $\chi(p)$ functions into a finite basis set:
\begin{align}
\label{Ansatz}
  \chi(p) =\sum_{m=1}^N a_m \phi^{(m)}(p) .  
\end{align}
We will use the variational Ansatz proposed%
\footnote{To our knowledge this variational Ansatz was first used in Ref.~\cite{Jeon:1995zm}, but it is not actually presented there; we thank Larry Yaffe for private communications on this point.}
in Ref.~\cite{Jeon:1995zm,Arnold:2000dr}:
\begin{equation}
\label{LGYbasis}
    \phi^{(m)}=\frac{p(p/T)^m}{(1+p/T)^{N-1}}, \hspace{0.3cm} m=1,...,N \,.
\end{equation}
The subspace spanned by this basis strictly increases as $N$ is increased.

\section{Collision operators for an NLO evaluation}
\label{sec:NLO}

At vanishing chemical potential, the perturbative framework suggests that the QGP predominantly comprises quasiparticle excitations with momenta of the order of the temperature.  
Their dispersion and occupancy are corrected only at $\OO(g^2)$, which for an $\OO(g)$-subleading calculation means that we can ignore corrections both to the lowest-order entropy density $s$ and to the lowest-order particle dispersion relations (except for splitting, see below).
Therefore NLO corrections will only be needed in the collision integral, which we turn to next.
The LO calculation \cite{Arnold:2003zc} requires two types of scattering processes, the $\twotwo$ scatterings with all hard ($\OO(T)$) external particles and $\onetwo$ effective splitting processes between hard participants. 

At NLO, all the scatterings and corrections to be added are the ones suppressed by a single power of $g$.  As shown in detail in \cite{Ghiglieri:2018dib}, there are only a
few such $\OO(g)$ subleading effects, which can be described as follows. The rate of soft
$\twotwo$ scattering is modified to be considered as an
additional momentum-diffusion coefficient $\delta\hat{q}$.  This along with an $\OO(g)$ correction to the medium corrections to dispersion provides an $\OO(g)$ shift in the $\onetwo$ splitting rate.  Next, the $\onetwo$ splitting rate must be corrected wherever one participant of the process is "soft", $p \sim gT$, or the opening angle is less collinear.  Finally, the numerical implementation of the LO scattering kernel \cite{Arnold:2003zc} already resums a small amount of these NLO effects, requiring a subtraction to $\delta\hat{q}$ and the longitudinal momentum diffusion, $\delta \hat{q}_L$. 
This section will briefly review the collision operator at LO and NLO at vanishing chemical potential.

\subsection{Collision Integrals at LO}
\label{subsec:LOcollision}

The leading-order collision operator encodes the contribution of tree-level $\twotwo$
scattering processes, with Hard-Loop resummed
propagators in the soft-sensitive channels,
as well as collinear, effective $\onetwo$ processes resumming the effect
of splitting caused by any number of soft scatterings.
Both processes contribute to order $g^4T$ to the collision operator; a subset of $C_a^{\twotwo}[f]$
is logarithmically enhanced, $g^4T\ln(1/g)$, due to the sensitivity
to the soft scale $gT$.
$C_a^{\twotwo}[f]$ and $C_a^{\onetwo}[f]$ are described in detail
in \cite{Arnold:2002zm,Arnold:2003zc}. Therefore, the collision term has the form:
\begin{equation}
    \mathcal{C}_a[f] = \mathcal{C}_a^{2\leftrightarrow 2}[f] + \mathcal{C}_a^{1\leftrightarrow 2}[f]
    \label{collop}
\end{equation}

\begin{figure}
    \centering
    \includegraphics[scale=0.55]{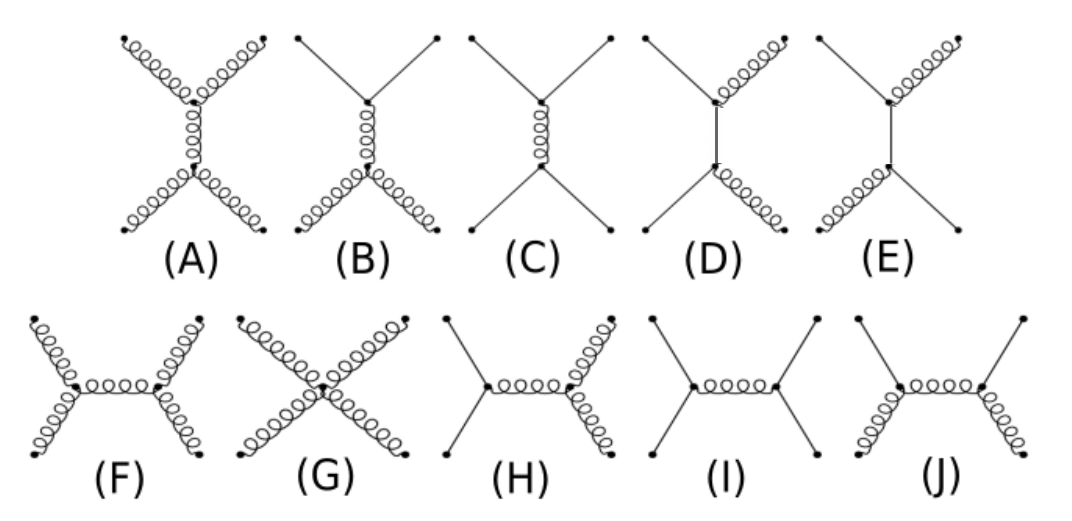}
    \caption{The diagrams relevant to leading-order scattering in QCD, showing our labeling conventions.}
    \label{diag}
\end{figure}

The matrix elements that contribute to $2 \leftrightarrow 2$ processes at leading order come from the diagrams shown in Fig.~\ref{diag}. In terms of the collision operator, these contributions can be written as:
\begin{align}
    \nonumber
    \CC^a[f](\vec{p}) =\frac{1}{2} \sum_{bcd} \int_{\vec{k},\vec{p},\vec{k'}} &
    \frac{|\mathcal{M}_{abcd}(P,K,P',K')|^2}{2p^0\, 2k^0\, 2p'{}^0\, 2k'{}^0}
    (2\pi)^4 \delta^4(P+K-P'-K') \\
    \nonumber
    &\times \Big\{ f^a(\vec{p})f^b(\vec{k})[1\pm f^c(\vec{p'})][1\pm f^d(\vec{k'})] \\
    & \phantom{\times \Big\{ } {} -  f^c(\vec{p'})f^d(\vec{k'})[1\pm f^a(\vec{p})][1\pm f^b(\vec{k})]\Big\}
    \label{collop22}
\end{align}
where the incoming/outgoing momenta $\vec p,\vec k$ and $\vec p',\vec k'$ are all on shell, $p^0=p$. The squared matrix element 
$\left|\mathcal{M}_{cd}^{ab} \right|$ is hard thermal loop (HTL) resummed
\cite{Braaten:1989mz,Frenkel:1989br}
and is summed over all spin polarizations and colors.  We write $\int_{\vec{k}} = \int \frac{d^3\vec{k}}{(2\pi)^3}$ to simplify the notation. 

Diagrams (D) and (E) are processes involving t (or u) channel fermion exchange.
These processes have a logarithmic infrared divergence when computed using free propagators.
To cut off this divergence, one can include the retarded HTL fermionic thermal self-energy, as was originally done in Ref.~\cite{Arnold:2003zc}.
It was later shown in Ref.~\cite{Kurkela:2018oqw} that there is a simpler treatment which is also correct at the leading order.
The most important feature of the HTL self-energy insertion is that it correctly reproduces the total scattering rate for this process; since the process is nearly isotropic, the detailed angle dependence is not relevant.
A mass-like insertion, with the correctly chosen coefficient, reproduces this rate and is much simpler to use in calculation.
Specifically for a soft fermion exchange in the t or u channel, one can simply use:
\begin{equation}
\label{quarkscreening}
    \frac{u-s}{t} \rightarrow \frac{u-s}{t}\frac{q^2}{q^2 + \xi_q^2 m^2_{\infty,q}}
\end{equation}
where $q$ is the exchanged momentum, $\xi_q = e/2 \simeq 1.359$, and $m^2_{\infty,q}$ is the medium-induced effective quark mass-squared for quarks, given by
\begin{align}
\label{thermalquarkmass}
 m^2_{\infty,q}  & = 2g^2\int \frac{d^3 \bf{p}}{2p^0(2\pi^3)}[2\cf f_g(\textbf{p}) + \cf (f_q(\textbf{p})+ f_{\bar{q}}(\textbf{p}) )]
 \nonumber \\
 & = \frac{g^2 \cf}{4} \left( T^2 + \frac{\mu^2}{\pi^2} \right) = \frac{g^2}{3} \left( T^2 + \frac{\mu^2}{\pi^2} \right).
\end{align}
We have allowed ourselves to make this simplification -- also because the cumulative effect of these diagrams is rather small and is exponentially suppressed $\propto e^{-\mu/T}$ in the large $\mu/T$ region.

On the other hand, diagrams (A),(B), and (C) represent processes involving t or u-channel gauge boson exchange that require the inclusion of the thermal
gauge boson self-energy on the internal propagator to cut off the infrared sensitivity of these processes. The matrix element of interest is $(s^2+u^2)/t^2$ which can be written as:
\begin{equation}
\label{gluonprop}
    \frac{1}{8}D_{\mu\alpha}^{\mathrm{\mathrm{ret}}}D_{\nu\beta}^{\mathrm{adv}}\Trace(\slashed{p}\gamma^\mu \slashed{k}\gamma^\nu)\Trace(\slashed{p'}\gamma^\alpha\slashed{k'}\gamma^\beta)
\end{equation}
where $D_{\mu\alpha}^{\mathrm{ret}}$ and $D_{\mu\alpha}^{\mathrm{adv}}$, are the retarded and advanced boson propagators, respectively.
Following the same notation as \cite{Moore:2001fga}, the plasma frame frequency and momentum
carried by the gauge boson propagator are denoted as $\omega$ and $q$. The retarded gauge boson propagator can be conveniently expressed in Coulomb gauge as:
\begin{align}
\nonumber
    &D_{00}^{\mathrm{ret}}(\omega,\textbf{q})=\frac{-1}{q^2 - \PiL^{\mathrm{ret}}(\omega,\textbf{q})}\\
    \label{coulombgauge}
    &D_{ij}^{\mathrm{ret}}(\omega,\textbf{q})=\frac{\delta_{ij}- \hat{\textbf{q}}_i\hat{\textbf{q}}_j}{q^2 - \omega^2 +\PiT^{\mathrm{ret}}(\omega,\textbf{p})}\\
    &D_{0i}^{\mathrm{ret}}(\omega,\textbf{q})=D_{i0}^{\mathrm{ret}}(\omega,\textbf{q})=0 .
\nonumber
\end{align}
The equilibrium transverse and longitudinal gauge boson self-energies in the HTL approximation are \cite{Klimov1,Weldon1}
\begin{align}
    \PiT^{\mathrm{ret}}(\omega,\q) & =  \mD^2
	\left\{
	    \frac{\omega^2}{2q^2} 
	+
	    \frac{\omega \, ( q^2 {-} \omega^2 )}{4 q^3}
	    \left[
		\ln \left( \frac{q+\omega}{q-\omega}\right) - i \, \pi
	    \right]
	\right\}
	\, , 
\label{eq:PiT} \\
    \Pi^{\mathrm{ret}}(\omega,\q) & =  \mD^2
	\left\{
	    1
	    -
	    \frac{\omega}{2q}
	    \left[
		\ln \left(\frac{q+\omega}{q-\omega}\right) - i \, \pi
	    \right]
	\right\} \, ,
\label{eq:PiL}
\end {align}
where we have assumed $|\omega| < q$,
which is the only case of relevance. 
Here $\mD^2$ is the squared Debye mass, which we will give for the case of nonzero $\mu,T$ in \Eq{mDebye}.
The advanced propagators are the complex conjugates of the retarded ones.

The remaining diagrams and the interference terms are all finite and can be computed directly. One should refer to \cite{Arnold:2003zc} for more details on these calculations.

The second type of contribution to leading order comes from $1 \leftrightarrow 2$ nearly collinear processes. These can be approximated as strictly collinear, which only produces relative changes suppressed by at least one power of g.  Corrections at this order will be treated in the next subsection.
The leading-order computation proceeds as follows
\cite{Arnold:2002zm}:
\begin{align}
    \nonumber
    \mathcal{C}_a^{1\leftrightarrow 2}[f](\vec{p}) = &\frac{(2\pi)^3}{2|\vec{p} |^2\nu_a}\sum_{bc}\int_0^{\infty} dp' dk \delta(|\textbf{p}|-p'-k')\gamma^a_{bc}(\textbf{p};p'\hat\p,k'\hat\p)\,\\
    \nonumber
    &\left\{f^a(\textbf{p})[1\pm f^b(p'\hat\p)][1 \pm f^c(k\hat\p)]-f^b(p'\hat\p) f^c(k\hat\p)[1 \pm f^a(\textbf{p})]\right\}\,\\
    \nonumber
    +&\frac{(2\pi)^3}{2|\vec{p} |^2\nu_a}\sum_{bc}\int_0^\infty dk dp' \delta(|\textbf{p}|+k-p')\gamma^c_{ab}(p'\hat\p;\textbf{p},k\hat\p)\,\\
    &\left\{f^a(\textbf{p})f^b(k\hat\p)[1 \pm f^c(p'\hat\p)]-f^c(p'\hat\p) [1 \pm f^a(\hat\p)][1 \pm f^b(k\hat\p)]\right\},
    \label{collop21}
\end{align}
where $\gamma_{bc}^a$ represents the splitting rate, and is given by \cite{Arnold:2002zm}:
\begin{equation}
\gamma_{bc}^a(p;p-k,k)= \frac{g^2d_{R_b}C_{R_b}}{64\pi^4}
\left\{ \begin{array}{ll}
\frac{p^4+k^4+(p-k)^4}{p^3k^3(p-k)^3} & g \leftrightarrow gg\\
\frac{p^2+(p-k)^2}{p^2k^3(p-k)^2} & q \leftrightarrow qg \\
\frac{p^2+(p-k)^2}{p^3k^2(p-k)^2} & g \leftrightarrow q\Bar{q} \\
\end{array} \right\} \times
\int \frac{d\textbf{h}^2}{(2\pi)^2}2\bf h\cdot \Re \textbf{F}_b,
\label{split}
\end{equation}
where $\bf h= \textbf{p} \times \textbf{k}$ is a transverse, two-dimensional vector related to the transverse momentum picked up during the splitting process. And $\textbf{F}_b$ can be obtained through the following integral equation:
\begin{align}
\label{split_int}
    2\bf= & i\delta E(\textbf{h})\textbf{F}_b(\textbf{h}) + \int \frac{dq_\perp^2}{(2\pi)^2} \Bar{C}(q_\perp) \{\left(C_{R_b}-C_A/2\right)[\textbf{F}_b(\textbf{h})-\textbf{F}_b(\textbf{F}-k\textbf{q}_\perp)]\\ \nonumber
    & C_A/2[\textbf{F}_b(\textbf{h})-\textbf{F}_b(\textbf{h}+p\textbf{q}_\perp)] + C_A/2[\textbf{F}_b(\textbf{h})-\textbf{F}_b(\textbf{h}-(p-k)\textbf{q}_\perp)]\}.
\end{align}

For the case of $g \leftrightarrow q \bar{q}$, $C_{R_b} - C_A/2 = \cf - C_A/2$ multiplies the term with $\textbf{F}_b(\textbf{h}-p\textbf{q}_\perp)$
rather than $\textbf{F}_b(\textbf{h}-k\textbf{q}_\perp)$. Here $\Bar{C}(q_\perp)$ represents the  leading-order transverse scattering kernel:
\begin{equation}
\label{scattkernel}
    \Bar{C}(q_\perp)=\frac{g^2T\mD^2}{\textbf{q}_\perp^2(\textbf{q}_\perp^2+\mD^2)},
\end{equation}
and $\delta E(\textbf{h})$ is the energy difference between initial and final collinear particles, expanded to first nontrivial order in thermal masses and transverse momenta:
\begin{equation}
\label{deltaEeq}
    \delta E(\textbf{h}) = \frac{h^2}{2pk(p-k)}+\frac{m^2_{\infty,c}}{2k}+\frac{m^2_{\infty,b}}{2(p-k)} - \frac{m^2_{\infty,a}}{2p}.
\end{equation}
We encountered $m^2_{\infty,q}$ already in \Eq{thermalquarkmass}.
For gluons, $m^2_{\infty,g} = \mD^2/2$.

\subsection{Next-to-Leading Order corrections}
\label{subsec:nlo}

We now focus on the next-to-leading order corrections.
Treating kinetic theory to NLO was pioneered in
Ref.~\cite{Ghiglieri:2015ala,Ghiglieri:2015zma} and applied to QCD shear viscosity in Ref.~\cite{Ghiglieri:2018dib}.
In this work, the authors observed that the analytically complex parts of the collision operator do not receive $\OO(g)$ corrections, which instead only modify collinear exchange, incorporated by changing $\gamma^a_{bc}$,
and soft exchange, which can be modeled as momentum-diffusion and identity-changing processes.
We briefly review the extended treatment that captures these $O(g)$ corrections, and then we summarize it and present an explicit form of the collision operator at NLO.

\subsubsection{Momentum diffusion}
\label{subsubsec:softglue}

\begin{figure}[tbh]
\centerline{\includegraphics[width=0.95\textwidth]{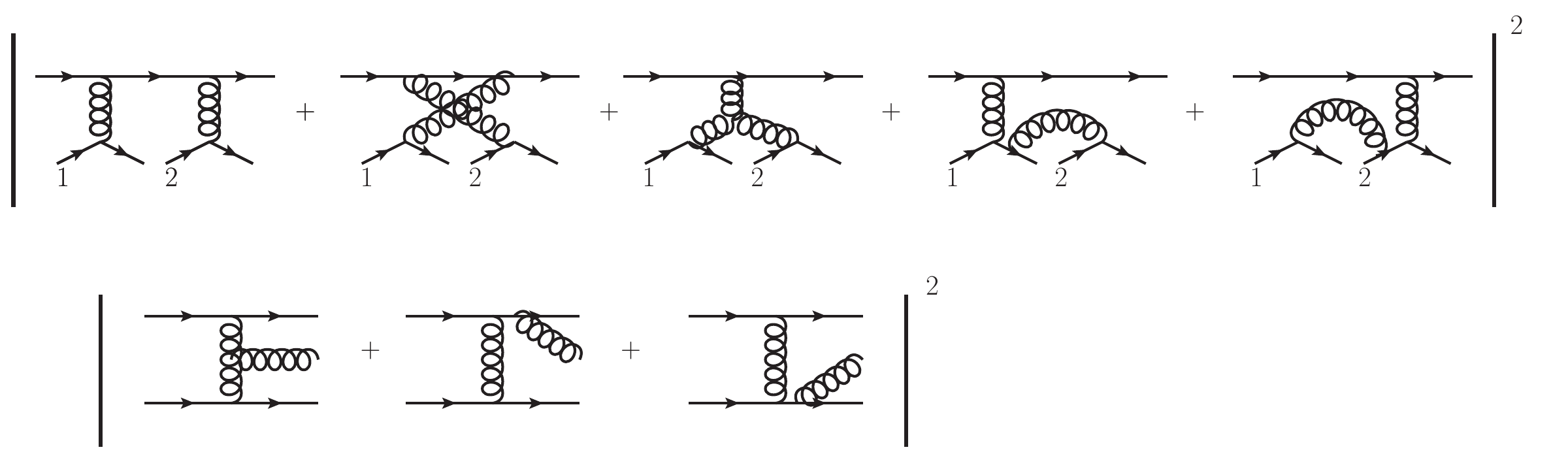}}
\caption{\label{fig:NLOscatt} Some scattering effects which affect momentum diffusion at NLO.}
\end{figure}

The small momentum exchange part of diagrams (A), (B), (C) can be well approximated by momentum diffusion for the particles on each line.
Beyond leading order, there are additional small-momentum-exchange effects which are only suppressed by $\OO(g)$ relative to LO, which must be incorporated.
One such effect is the possibility for the gluon in diagrams (A), (B) to carry a soft momentum.
Other effects involve the interference between multiple soft scatterings, soft gluon emission during scattering, and nonabelian interactions between scattering processes, such as those depicted in Figure~\ref{fig:NLOscatt}.

Because the momentum exchanges involved in these processes are small (that is the reason they can occur at $\OO(g)$ and not $\OO(g^2)$), these terms can be adequately described in terms of Langevin dynamics, that is, in terms of momentum drag and longitudinal and transverse momentum diffusion.
This is analyzed in the context of heavy quark dyamics in Ref.~\cite{Moore:2004tg} and jet energy loss in Ref.~\cite{Ghiglieri:2015zma}, and applied to shear viscosity in Ref.~\cite{Ghiglieri:2018dib}.
As shown there, the effective collision term due to soft scatterings, cut off at a scale $\mu_\perp$, is
 \begin{equation}
    \label{FPall}
	(C_{\mathrm{diff}}^{\twotwo} f_1)^a(\p) =
    -\frac{1}{2} \frac{\partial}{\partial p^i} \,
    \hat q^{ij}_a \,  f_0^a(p) (1 {\pm} f_0^a(p))
    \frac{\partial f_1^a(\p)}{\partial p^j}  + \mbox{gain-terms} \, ,
 \end{equation}
where
\begin{equation}
  \label{Eq:qhatLT}
  \hat q^{ij}_a = \hat q_{a,L}  \hat p^i \hat p^j
  +  \tfrac{1}{2} \hat q_a \left(\delta^{ij} - \hat p^i \hat p^j \right)
\end{equation}
represents the momentum diffusion parallel and perpendicular to the particle's momentum through $\hat q_{L}$ and $\hat q$ respectively.

Contracting with the departure from equilibrium and integrating over $\p$, the contribution to the inner product of \Eq{Qdef} takes the form
\st
\label{lossgain}
 (f_1, C^{2 \rightarrow 2}_{\mathrm{diff}} f_1)  \equiv
 \left. (f_1, C^{2 \rightarrow 2}_{\mathrm{diff}} f_1)\right|_{\rm loss}
 +
 \left. (f_1, C^{2 \rightarrow 2}_{\mathrm{diff}} f_1)\right|_{\rm gain} \, ,
 \stp
 where
 \st
 \label{FPloss}
 \left. (f_1, C^{2 \rightarrow 2}_{\scriptscriptstyle\rm diff} f_1) \right|_{\rm loss}
= \frac{1}{2} \beta^3 \sum_{a} \nu_a \,
\int_\p  f_0^a(p)(1 \pm f_0^a(p)) \,  \hat q^{ij}_{a} \, \frac{\partial f_1^a(\p) }{\partial p^i} \frac{\partial f_1^a(\p) }{\partial p^j } \, ,
\stp

The gain collision term takes the form
\begin{align}
   \mbox{gain-terms}  = {} &
   \frac{1}{2} \frac{\partial}{\partial p^i } \left(
   f_0^a(p) (1 {\pm} f_0^a(p))  \, \sum_{b} \nu_b \!\int_\k
   \A^{ij}_{ab} (\hat \p \cdot \hat \k) \, f_0^b(k) (1 \pm f_0^{b}(k))
   \frac{\partial f_1^b(\k)}{ \partial k^j } \right)
\nonumber \\
\label{gain}
\left. (f_1, C^{2 \rightarrow 2}_{\mathrm{diff}} f_1)\right|_{\rm gain} 
= {} &  -\frac{\beta^3}{2}  \sum_{ab} \nu_a \nu_b \!\int_{\p\k} \!\!
f_0^a(p)(1 \pm f_{0}^a(p)) f^b_0(k) (1 \pm f_{0}^b(k))
\nonumber \\ & \hspace{7em} {} \times 
 \A_{ab}^{ij} (\hat\p\cdot \hat\k) \frac{\partial f_1^a(\p) }{\partial p^i}
  \frac{\partial f_1^b(\k)}{\partial k^j},
\end{align}
where the angular function is
\st
\label{eq:qhatkernel}
\A_{ab}^{ij}(\hat\p\cdot \hat\k)
=  \frac{g^4 C_{R_a} C_{R_b}}{d_A} \int \frac{d^4Q}{(2\pi)^4}   |G_{\mu\nu}^R(Q) v_{\p}^\mu v_\k^{\nu}|^2  \,  2\pi \delta(v_\p\cdot Q) 2\pi\delta(v_\k \cdot Q) q^i q^j \, ,
\stp
which is related to $\hat q^{ij}_a$ through
\st
 \label{qhatintegral}
 \hat q^{ij}_a =  \sum_b \nu_b \int_\k f_0^b(k) (1 \pm f_0^{b}(k)) \, \A_{ab}^{ij}(\hat \p \cdot\hat \k)  \,.
\stp
Beyond leading order it turns out to be much easier to determine $\hat q^{ij}_a$ than $\A^{ab}_{ij}(\hat \p \cdot \hat \k)$,
since light-cone techniques work for the former but not the latter.
Therefore, in practice we will not be able to compute this gain term.
Since our best estimates based on leading-order analyses indicate that angular cancellations make this term small, we will instead drop it.
This is why we generally refer to our results as ``almost next-to-leading order.''

Examining the expression for
$\left. (f_1, C_{\mathrm{diff}}^{\twotwo}f_1) \right|_{\rm loss}$,
it becomes evident that it involves a single light-like vector denoted as $v_\p$. 
Using the causality and KMS properties of such light-like correlators~\cite{CaronHuot:2008ni}, these leading-order
soft contributions to $\hat{q}$ and $\hat{q}_L$  can be evaluated
in closed form \cite{Aurenche:2002pd,Ghiglieri:2015zma,Ghiglieri:2015ala},
\begin{equation}
	\hat {q}^a\bigg\vert_\mathrm{soft}=\frac{g^2C_{R_a} T
	\md^2}{2\pi}\ln\frac{\mu_\perp}{\md},	\qquad
	\hat {q}_{\sss L}^a\bigg\vert_\mathrm{soft}=\frac{g^2C_{R_a} T
	\md^2}{4\pi}\ln\frac{\sqrt{2}\mu_\perp}{\md}.
	\label{qhatql}
\end{equation}
In these expressions, $gT\ll\mu_\perp\ll T$ is a cutoff on the $q_\perp \equiv \sqrt{q^2 - \omega^2}$
integration, which separates the soft from the hard scale. In the region where $\omega,q\gtrsim T$, and the bare matrix elements can be used to compute the hard contribution to the collision operator $C^{\twotwo}$, there is a cancellation on this cutoff dependence. Since light-like separated points are effectively causally disconnected
as far as the soft gauge fields are concerned,  $\hat{q}$ and $\hat{q}_L$ have a simple form.
Using the explicit form of the angular dependence of
$f_1^a(\p) = \beta^2 \chi_{ij}(\p) X_{ij}$,
straightforward analysis shows that the loss term  reduces to
\begin{multline}
    \left. \Big( \chi_{ij}, C^\twotwo_{\mathrm{lossshort}} \, \chi_{ij} \Big) \right|_{\rm loss} =
 	 \frac{\beta^3}{2}
 	  \sum_a \nu_a\int_\p f^a_0(p)
 	 (1\pm f_0^a(p)) \times
         \\
 	\left[(\chi^{a}(p)')^2  \hat {q}^a_{\sss L}\bigg\vert_\mathrm{soft}
 	+
 	\frac{\ell(\ell+1)\chi^a(p)^2}{2p^2}
        \hat {q}^a\bigg\vert_\mathrm{soft}\right] \, ,
\label{diffterms}
\end{multline}
which is the most useful form for evaluating the transport coefficients
numerically.

The next-to-leading order value of $\hat{q}^a$ has been computed \cite{CaronHuot:2008ni}.
This result can be simply included by adding it to the above results \cite{Ghiglieri:2018dib}:
  \begin{equation}
    \delta\hat {q}^a=\frac{g^4C_{R_a} C_A \md T^2 }{32\pi^2}
    \left(3\pi^2+10-4\ln 2\right) \,.
    \label{qhatnlo}
  \end{equation}
the shift included in the soft gluon exchange diffusion rate gives
 \begin{equation}
 	\Big( \chi_{ij}, C^{\delta \Hat{q}} \, \chi_{ij} \Big) =
 	 \frac{\beta^3}{2}\sum_a\nu_a\,\delta \Hat{q}^a
 	 \int_\p   f^a_0(p)
 	[1\pm f_0^a(p)]
 	 \frac{\ell(\ell+1)\chi^a(p)^2}{2p^2} .
 	\label{cnloqhat}
 \end{equation}

As discussed in Ref.~\cite{Ghiglieri:2018dib}, the precise procedure described in the last subsection for including HTL effects in the leading-order calculation also introduces some NLO effects in that calculation.
In particular, the phase space and statistical functions for scattering from soft gluons incorporate some of the $\OO(g)$ effects which we have just described.
Therefore, the leading-order procedure we use will not in fact reproduce \Eq{qhatql}, but will reproduce that expression plus some $\OO(g)$ additional terms.
These contributions must then be subtracted from the $\OO(g)$ contribution we include here, in order to avoid overcounting such effects.
Fortunately, since the effects arise from scattering off of gluons, they are unmodified by the inclusion of quark chemical potentials, and we can take over the analysis of
Ref.~\cite{Ghiglieri:2018dib}.
The resulting counterterm can be written as
 \begin{eqnarray}
	\Big( \chi_{ij}, C^\twotwo_{\OO(g)\,\mathrm{finite}\,g} \, \chi_{ij} \Big) &=&
	\frac{d_AC_Ag^4 \md}{32\pi^5 T}\sum_a T_{R_a}
	  \int_{0}^\infty dp\, p^2\,f^a_0(p)[1\pm f_0^a(p)]\nn\\
	  &&\hspace{-2cm}
	  \times\bigg\{4.2695\big[(\chi^{a}(p)')^2+(\chi^{g}(0)')^2\big]
	  +7.1769\frac{\ell(\ell+1)}{2p^2}\big[(\chi^{a}(p))^2+(p\chi^{g}(0)')^2\big]\nn\\
	  &&\hspace{5cm}+
	  18.0669\,\delta_{\ell2}[\chi^{g}(0)']^2
	  \bigg\},
	\label{finalog}
\end{eqnarray}
and the soft fermion exchange counterterm is given by
\begin{align}
\Big( \chi_{ij}, C^\twotwo_{\OO(g)\,\mathrm{finite}\,q} \, \chi_{ij} \Big) =& \,\frac{ d_F \cf^2\nf g^4
m_\infty  }{32\pi^5T^2}\,9.95268\,
\int_{0}^\infty dk\,k\,
 f^q_0(k)	[1+ f_0^g(k)]\nn\\
 &\hspace{3.5cm}\times\big[(\chi^q(k)-\chi^g(k))^2+(\chi^q(0))^2\big]\,.
	\label{softktot}
\end{align}
One should refer to \cite{Ghiglieri:2018dib} for a detailed derivation of these counterterms.

\subsubsection{Collinear and semi-collinear contributions }
\label{subsubsec:semicollin}

The collinear splitting process, \Eq{collop21} to \Eq{deltaEeq}, involves the same momentum diffusion effects we just described.
And as we just described, the strength of momentum diffusion receives $\OO(g)$ corrections.
Furthermore, \Eq{deltaEeq} shows that collinear splitting is sensitive to thermal effective masses.
These also turn out to receive $\OO(g)$ corrections.
Therefore we need to analyze how these two corrections impact the collinear splitting rate.
To conduct the analysis, we will follow Ref.~\cite{Aurenche:2002pd} and Fourier transform \Eq{collop21} from transverse-momentum $\q_\perp$ to transverse impact parameter $\bb_\perp$ space.
This not only diagonalizes the collision kernel $C(k_\perp)$, but also turns an integral equation into a differential equation.
Furthermore,  the source on the left-hand side becomes a boundary condition at $\textbf{b}=0$
and the desired final integral becomes a boundary
value of the ODE solution.
Hence, defining
\begin{equation}
\label{impactspace}
\textbf{F}(\bb) = \int \frac{d^2h}{(2\pi)^2} e^{i\bb \cdot \bf h}\textbf{F}(\textbf{h})\,,
\end{equation}
we have
\begin{equation}
\label{b_want}
\Re \int \frac{d^2h}{(2\pi)^2} 2  \cdot \textbf{F}(\textbf{q})
= \Im( 2 \nabla_{\bf b} \cdot \textbf{F}(0))\,,
\end{equation}
and \Eq{split_int} becomes
\begin{eqnarray}
\nn-2i\bfnabla \delta^2(\bb)&=&\frac{i}{2p\omega(p-\omega)}\left(p(p-\omega)m_{\infty\,\omega}^2+
p\,\omega\,m_{\infty\,p-\omega}^2-\omega(p-\omega)m_{\infty\,p}^2-\nabla_\bb^2\right)\textbf{F}(\bb)\\
&&+
\left(\cc'_R(\vert \omega\vert\,b)-\frac{\cc'_A(\vert \omega\vert\,b)}{2}
+\frac{\cc'_A(\vert p\vert\,b)}{2}+\frac{\cc'_A(\vert p-\omega\vert b)}{2}\right)\textbf{F}(\bb),
\label{bspace}
\end{eqnarray}
with
\begin{equation}
\label{eq:C(b)}
\cc_R'(\vert \omega \vert\,b) \equiv \int \frac{d^2\textbf{k}_\perp}{(2\pi)^2}
\Big( 1 - e^{i \omega \bb \cdot \textbf{k}_\perp} \Big) \cc_R(\textbf{k}_\perp)\,.
\end{equation}
As noted above, we need to incorporate the change in the collision operator,
\begin{equation}
\label{deltaC}
\cc'_{R\, ,{\rm LO+NLO}}(b) = \cc_R'(b) + \delta \cc_R'(b),
\end{equation}
where the $\q_\perp$-space correction of $\cc_R$ is the unintegrated version of $\delta \hat{q}^a$ detailed in \Eq{qhatnlo}.
The details appear in Appendix C of Ref.~\cite{Ghiglieri:2013gia}.
We must also incorporate a shift in the thermal masses:
\begin{align}
\label{mshift}
m^2_{\infty\,a,\rm LO+NLO} & = m^2_{\infty\,a} +\delta m^2_{\infty\,a},  &
\delta m^2_{\infty q} & = -g\cf\frac{T\mD}{2\pi}, &
\delta m^2_{\infty g} & = -gC_A\frac{T\mD}{2\pi}.
\end{align}
These shifts are incorporated into \Eq{bspace} by solving for their effects perturbatively, by treating $\textbf{F}(\bb)$
formally as an expansion in powers of $\delta m_\infty, \delta \cc$;
$\textbf{F}(\bb) = \textbf{F}_0(\bb)+\textbf{F}_1(\bb)+\ldots$  to first order. For details on the boundary conditions and the numerical evaluation of these
equations, one should refer to 
\cite{Ghiglieri:2013gia,Ghisoiu:2014mha,Ghiglieri:2014kma}

The above analysis considers $\OO(g)$ corrections to the regime in which the external states are collinear up to $\OO(g^2)$ corrections.
This allowed for simplifications in the analysis, in particular in expressing \Eq{deltaEeq} in terms of the leading mass and transverse-momentum corrections.
But this is not always the case.
Integrating over opening angles, \Eq{deltaEeq} breaks down if the external momenta $P,K$ obey $P\cdot K \sim gT^2$,
and this regime represents an $\OO(g)$ part of the splitting processes considered.
Similarly, for hard scattering, if the transfer momentum $Q$ becomes nearly lightlike, our treatment also breaks down, and this again represents an $\OO(g)$ effect.
Both regimes feature small angles, but not as small as in the collinear splitting processes; so we refer to them as \textsl{semi-collinear}.
The analysis of this kinematic region was carried out in Ref.~\cite{Ghiglieri:2015ala}, and we will follow the development there and in Ref.~\cite{Ghiglieri:2018dib}.

The good news is that, in treating this regime, the kinematic variable $\bf h$ becomes large, namely $\textbf{h} \gg gT^2$.
The collision operator then becomes subdominant, $\cc \ll \delta E$, and one can perturb in it to first order.
The lowest-order and first nontrivial-order solutions are
\begin{align}
  \label{Fseries}
  \textbf{F}_{b1}(\textbf{h}) & = 2\bh/i\delta E(\textbf{h}) \,, \\
  \nn
  \textbf{F}_{b2}(\textbf{h}) & = \frac{i}{\delta E(\textbf{h})}
  \int \frac{d^2 q_\perp}{(2\pi)^2}
  \bar C(q_\perp) \left\{ \left( C_{R_b} - \frac{C_A}{2} \right)
       [\textbf{F}_{b1}(\textbf{h}) - \textbf{F}_{b1}(\textbf{h}-k\textbf{q}_\perp)] \right.
       \\ & \quad
       \left. {}+ \frac{C_A}{2} [\textbf{F}_{b1}(\textbf{h}) - \textbf{F}_{b1}(\textbf{h}+p\textbf{q}_\perp)]
       + \frac{C_A}{2} [ \textbf{F}_{b1}(\textbf{h}) - \textbf{F}_{b1} (\textbf{q}-(p{-}k)\textbf{q}_\perp)]
       \right\} . \nn
  \end{align}
This is the same as treating the emission in the Bethe-Heitler limit.
However, at large $\textbf{h}$, $q^- = \delta E \sim gT \sim q_\perp$, which changes the kinematics of the scattering. Therefore,
$\bar C(q_\perp)$ must be recomputed.  A more accurate form for
$\bar C(q_\perp)$ in this regime is
\cite{Ghiglieri:2013gia,Ghiglieri:2015ala}
\begin{equation}
  \label{CLONLO}
  \bar C_{\textrm{NLO}}(q_\perp,\delta E) =
  \frac{g^2 T \md^2}
       {(q_\perp^2 + \delta E^2)(q_\perp^2 + \delta E^2 + \md^2)}
       + \frac{2 g^2 T \delta E^2}{q_\perp^2(q_\perp^2 + \delta E^2)}
       \,.
\end{equation}
Physically, this is a consequence of the
splitting being induced by an elastic scattering or by the
absorption of a soft on-shell particle, as observed in \cite{Ghiglieri:2018dib}. Therefore,
the already-computed LO $\onetwo$ contribution, corresponding to the small $\delta E$
limit, and the already-included LO $\twotwo$ contribution \cite{Ghiglieri:2013gia} need to be subtracted as:
\begin{equation}
  \label{CNLO}
  \delta \bar C(q_\perp,\delta E) =
  \frac{g^2 T \md^2}
       {(q_\perp^2 + \delta E^2)(q_\perp^2 + \delta E^2 + \md^2)}
       - \frac{g^2 T \md^2}{q_\perp^2(q_\perp^2+\md^2)}
       \,.
\end{equation}
The second term is the LO collinear form for $\bar C$ (the small
$\delta E$ limit of \Eq{CLONLO}).

The semi-collinear contribution is found by substituting \Eq{CNLO}
into \Eq{Fseries} and using it to evaluate \Eq{split_int} and finally
\Eq{collop21}, which results in
\begin{align}
  \nn	\gamma^{a}_{bc}\bigg\vert_\mathrm{semi}(p;p-k,k)
  & = \frac{g^2}{32\pi^4}\left\{
\begin{array}{cc}
	d_AC_A\frac{p^4+k^4+(p-k)^4}{p^3k^3(p-k)^3}& g\leftrightarrow gg\\
	d_F \cf \frac{p^2+(p-k)^2}{p^2(p-k)^2k^3} & q\leftrightarrow q g\\
	d_F \cf\frac{(p-k)^2+k^2}{(p-k)^2k^2p^3} & g\leftrightarrow q\bar q
\end{array}
\right.
\int\frac{d^2h}{(2\pi)^2}
\int\frac{d^2\qp}{(2\pi)^2} \delta\bar{C}(\qp,\delta E)\\
&\nn\hspace{-3cm}\times\left[\left(C_R-\frac{C_A}{2}\right)\left(\frac{\textbf{h}}{\delta E(\textbf{h})}-
\frac{\textbf{h}-k\textbf{q}_\perp}{\delta E(\textbf{h}-k\textbf{q}_\perp)}\right)^2+\frac{C_A}{2}\left(\frac{\textbf{h}}{\delta E(\textbf{h})}-
\frac{\textbf{h}+p\textbf{q}_\perp}{\delta E(\textbf{h}+p\textbf{q}_\perp)}\right)^2\right.\\
&\left.+\frac{C_A}{2}\left(\frac{\textbf{h}}{\delta E(\textbf{h})}-
\frac{\textbf{h}-(p-k)\textbf{q}_\perp}{\delta E(\textbf{h}-(p-k)\textbf{q}_\perp)}\right)^2\right],
	\label{seminew}
\end{align}
with $(C_R-C_A/2)$ appearing on the $\textbf{h} + p\textbf{q}_p$ term for
$g \leftrightarrow q\bar{q}$ processes.  This is then inserted into
\Eq{collop21}, resulting in
\begin{eqnarray}
	\nn
	\left(f_1,\mathcal{C}^\mathrm{semi}f_1\right)
	&\equiv&\frac{2\pi}{T^3}\sum_{abc}\int_0^\infty \!\! dp
        \int_0^p dk\,\gamma^a_{bc}\bigg\vert_\mathrm{semi} \!\! (p;p-k,k)
	f^a_0(p)[1\pm f^b_0(k)][1\pm f^c_0(p-k)]\\
	&&\times\left[f_1^a(\p)-f_1^b(k\hat\p)-
	f_1^c((p-k)\hat\p)\right]^2.
	\label{semicoll}
\end{eqnarray}

\subsubsection{Summary}
\label{subsubsec:summary}

In summary, the corrections to be added to the strict leading-order contributions
of the previous section produce a collision operator that is fully
NLO.  At NLO, the collision term will be given as
\begin{equation}
	\label{fullnlo}
    \Big( f_1, C_{NLO} \, f_1 \Big)
    {}=
    \Big( f_1, C_{LO}\, f_1 \Big) +
    \Big( f_1, \delta C \, f_1 \Big) \,,
\end{equation}
where
\begin{align}
  \label{nlo1}
  \Big( f_1, \delta C \, f_1 \Big)
  \equiv &
  \Big( f_1, C^{\delta\Hat{q}} \, f_1 \Big)-
  \Big( f_1, C^\twotwo_{\OO(g)\,\mathrm{finite}} \, f_1 \Big)		  +
  \Big( f_1, C^{\semi} \, f_1 \Big)  +
  \Big( f_1, \delta C^{\onetwo} \, f_1 \Big).
\end{align}
Here, the first two terms are the contribution from the soft gluon exchange and its counterterm. The last two terms are the contribution from semi-collinear processes and the correction to the splitting rate, respectively. All of these corrections were carefully studied in \cite{Ghiglieri:2018dib}, and what was found is that the main difference between LO and NLO results arises from the soft gluon exchange. 

\section{From zero to nonzero chemical potential}
\label{sec:mu}

In this section, we delve into the differences between the description of the QGP at vanishing chemical potential, $\mu=0$, and at high densities, $\mu>0$.
Temperature, density, and screening enter the calculation, and we must make sure that we include them correctly each place they both arise.
The structure fits within the general effective kinetic framework of Ref.~\cite{Arnold:2002zm}, for which finite $\mu$ and $T$ case is a special case.

When introducing a chemical potential we must shift the equilibrium occupancies from their simple thermal values; we have already done this in \Eq{f0}.
This change propagates into the Boltzmann equation and the definition of the extremization problem, \Eq{Qdef}, in a straightforward way.
There is no need to change the basis of variational functions, but more terms may be needed because the most important momentum range is likely to be higher, $p \sim \mu$ rather than $p \sim T$.

The matrix elements depend on the chemical potential indirectly through the strength of screening effects.
As shown in Ref.~\cite{Braaten:1991gm}, the form of the HTLs are identical for high density as for high temperature.
The two effects simply combine in the Debye screening mass squared,
\begin{equation}
\label{mDebye}
    \md^2 = \frac{N_c}{3} g^2T^2 + \frac{\nf}{6} g^2\left(T^2 + \frac{3}{\pi^2}\mu^2\right).
\end{equation}
The last term represents the increase in the screening due to the quark abundance induced by the chemical potential.
The strength of quark screening is similarly enhanced; one uses the occupancies with chemical potentials in evaluating \Eq{thermalquarkmass}.
The resulting expression is already written in that equation.

In evaluating splitting rates, a pivotal role was played by the transverse scattering rate, described by the kernel $\bar{C}(\q_\perp)$.
Remarkably, \Eq{scattkernel} still holds at finite chemical potential, provided one uses the adjusted value of the Debye mass squared.
This is discussed in Ref.~\cite{Arnold:2002zm}.
Essentially, the shifted $\mD^2$ value accounts for the change in the number of scattering ``targets'' in the medium and therefore correctly adjusts the strength of medium scattering which induces the splitting.
Similarly, the calculation of $\delta E$ for splitting, \Eq{deltaEeq}, must use the form of the asymptotic masses which accounts for chemical potentials.

Similarly, at NLO we must correctly account for the strength of scattering in the medium, but generally the calculations presented so far remain true after one does so.
The light-cone methods which were used to derive \Eq{qhatql} and \Eq{qhatnlo} are based on integrating out the quarks and incorporating their effects in the HTLs or equivalently within the effective theory of EQCD.
This proceeds the same with $\mu$ as without, except that the value of $\mD^2$ must be adjusted using \Eq{mDebye}.
By similar reasoning, \Eq{mshift} and \Eq{CLONLO} apply as written, but using the corrected form of $\mD^2$.

Next, let us briefly analyze the formal convergence of perturbation theory.
At weak coupling $\als \ll 1$ and $\mu = 0$ hot QCD has three scales: the hard scale $\pi T$, the soft scale $gT$, and the ultrasoft scale $g^2 T$.
At leading order the scale $\pi T$ participates as the dominant propagating degrees of freedom, while the $gT$ scale is relevant as the momentum-exchange scale for soft scattering relevant in both $\twotwo$ and $\onetwo$ processes.
NLO corrections occur primarily because these $gT$ scale degrees of freedom have occupancies $f(q) \sim 1/g$ and loop corrections are only suppressed by a single power of $g$.
Therefore the structure of NLO $\OO(g)$ corrections relative to LO is often as a ratio $g^2 T/\mD$.
Including color and other numerical factors, these corrections often prove to be quite large.

The inclusion of a chemical potential with $\mu > \pi T$ raises the hard scale from $\pi T$ to $\mu$, and it increases the Debye scale as one sees in \Eq{mDebye}:
$\mD \sim g \mu$ rather than $gT$.
Therefore the separation between the soft and ultrasoft scales is increased, and $\OO(g)$ corrections are expected to scale as $g^2 T / g\mu$ instead of $g^2 T/gT$.
This underlies the expectation that NLO effects should be milder in the large-$\mu/T$ case.

In practice, the largest NLO effect is the shift in the transverse momentum diffusion coefficient, \Eq{qhatnlo}.
This effect is linear in $\mD$, while the leading-order part is proportional to $\mD^2$.
The ratio $\delta \hat{q} / \hat{q} \sim g^4 \mD T^2/g^2 \mD^2 T$ is therefore of order $g^2 T/\mD$, as anticipated above.
For parametrically large $\mu/T$, one indeed finds an improvement in the NLO convergence by a single factor of $T/\mu$.
In the next section we will see to what extent this really improves the convergence in practice.

\section{Results}
\label{sec:results}
In the last sections, we have presented the ingredients necessary for the computation of shear viscosity at leading order, discussed the relevant next-to-leading-order corrections, and presented a good basis set for these calculations.
Using this computational apparatus, we compute the shear viscosity as a function of $\mu/T$ at both leading order and (almost) next-to-leading order.

\begin{figure}[tbh]
    \centering
    \includegraphics[width=0.73\linewidth]{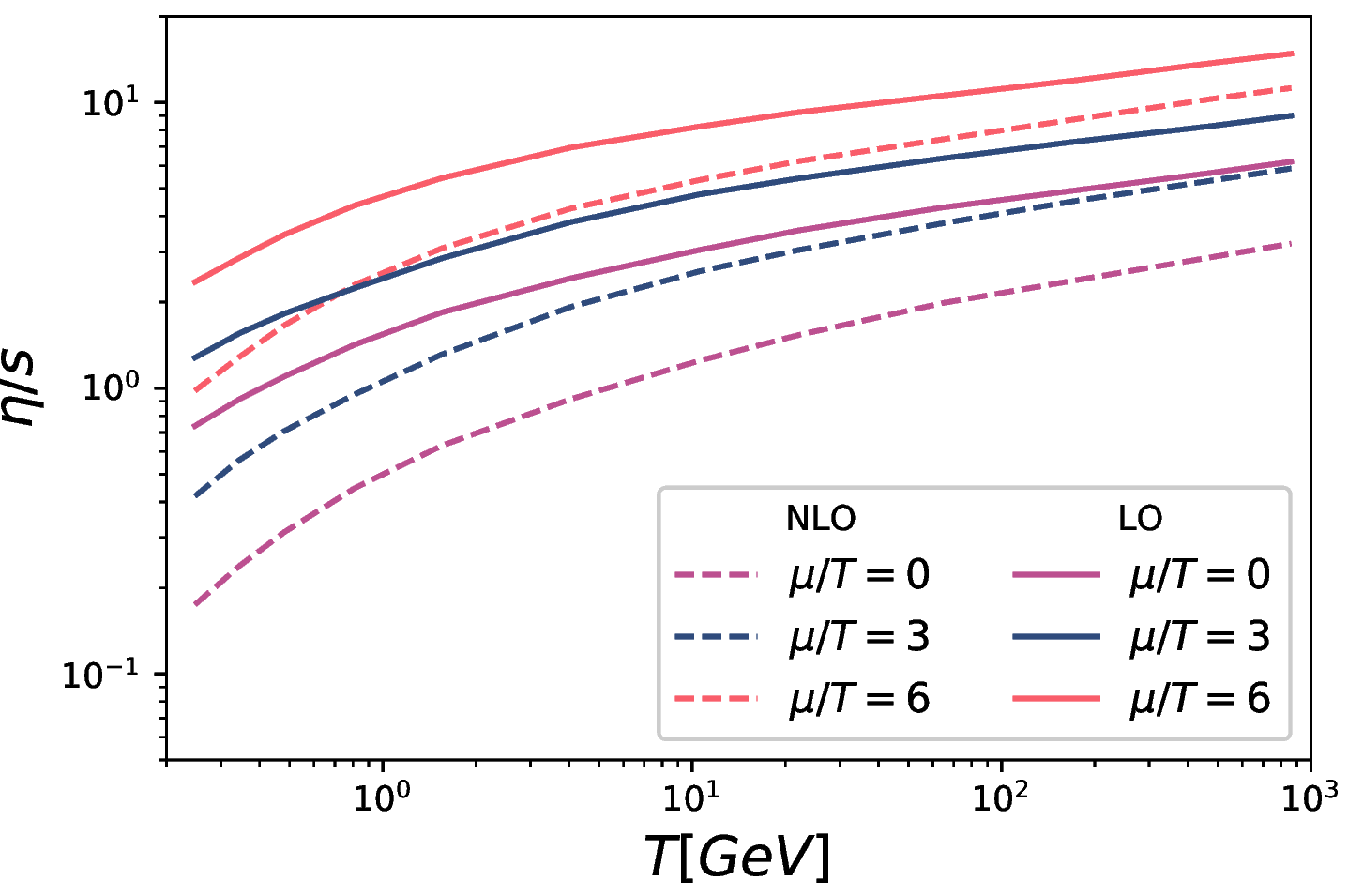}
    \caption{$\eta/s$ as a function of the temperature for $\mu=0,3,6$. The dashed lines represent NLO order results, while continuous lines are LO results. Both were calculated using the renormalization point $\muEQCD=2.7T$}
    \label{visc_ent}
\end{figure}

\begin{figure}[tb]
    \centering
    \includegraphics[width=0.75\linewidth]{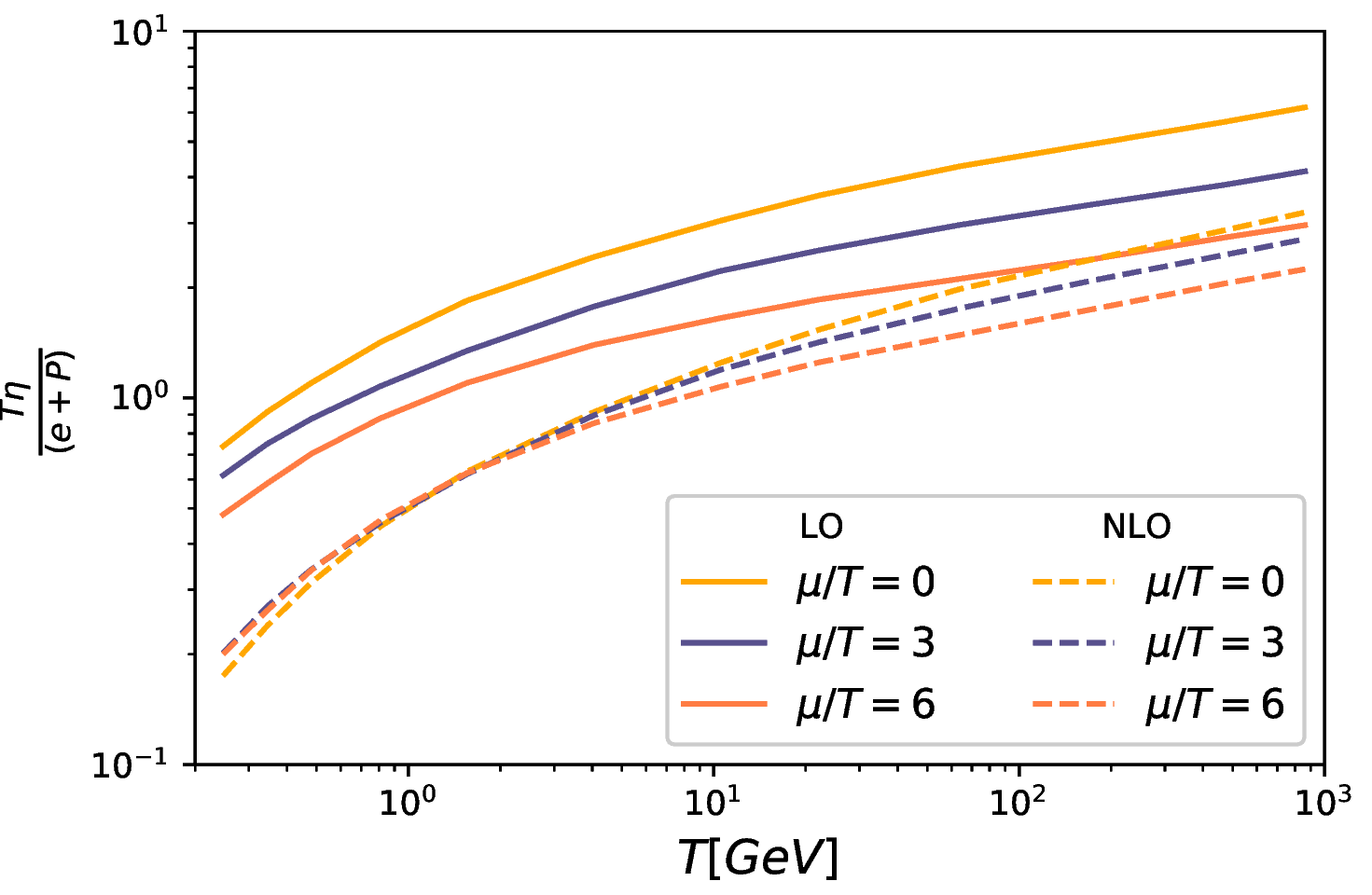}
    \caption{Kinematic shear viscosity as a function of the temperature for $\mu=0,3,6$. The dashed lines represent NLO order results, while continuous lines are LO results. Both were calculated using the coupling $\muEQCD=2.7T$}
    \label{kin_visc}
\end{figure}

In figure \ref{visc_ent} we plot the shear-viscosity to entropy density ratio $\eta/s$ for LO and NLO, and in figure \ref{kin_visc} we
show the kinematic viscosity, that is, the ratio of the shear viscosity to the enthalpy density, $\eta T /(e + P )$.%
\footnote{Raw data for all plots are included in the ancillary materials.}
Note that the chemical potential $\mu$ we refer to is that for a quark, and we treat it to be the same for all quark species.
The baryon number chemical potential is then 3 times larger.
We have used the EQCD-inspired scale setting by choosing the $\overline{\mathrm{MS}}$ renormalization point to be
$\muEQCD=2.7T$, following the treatment from \cite{Ghiglieri:2018dib}, and we run the coupling using the 2-loop beta function and the value $\als = 0.118$ at the $Z$-pole.
We assume that when $\muEQCD$ reaches a quark mass, one crosses the quark-mass threshold and the quark becomes active both in the medium and in the running of the coupling constant.
So for temperatures below $2.7T=m_b$ we use $\nf=4$ and for temperatures above $2.7T=m_b$ we use $\nf=5$ (and similarly for the charm and top quarks).
The plots clearly show that the difference between LO and NLO is smaller at large $\mu$ than at $\mu=0$, particularly at high temperatures where the coupling is smaller and $\nf$ is larger.

\begin{figure}[tb]
    \centering
    \includegraphics[scale=0.23]{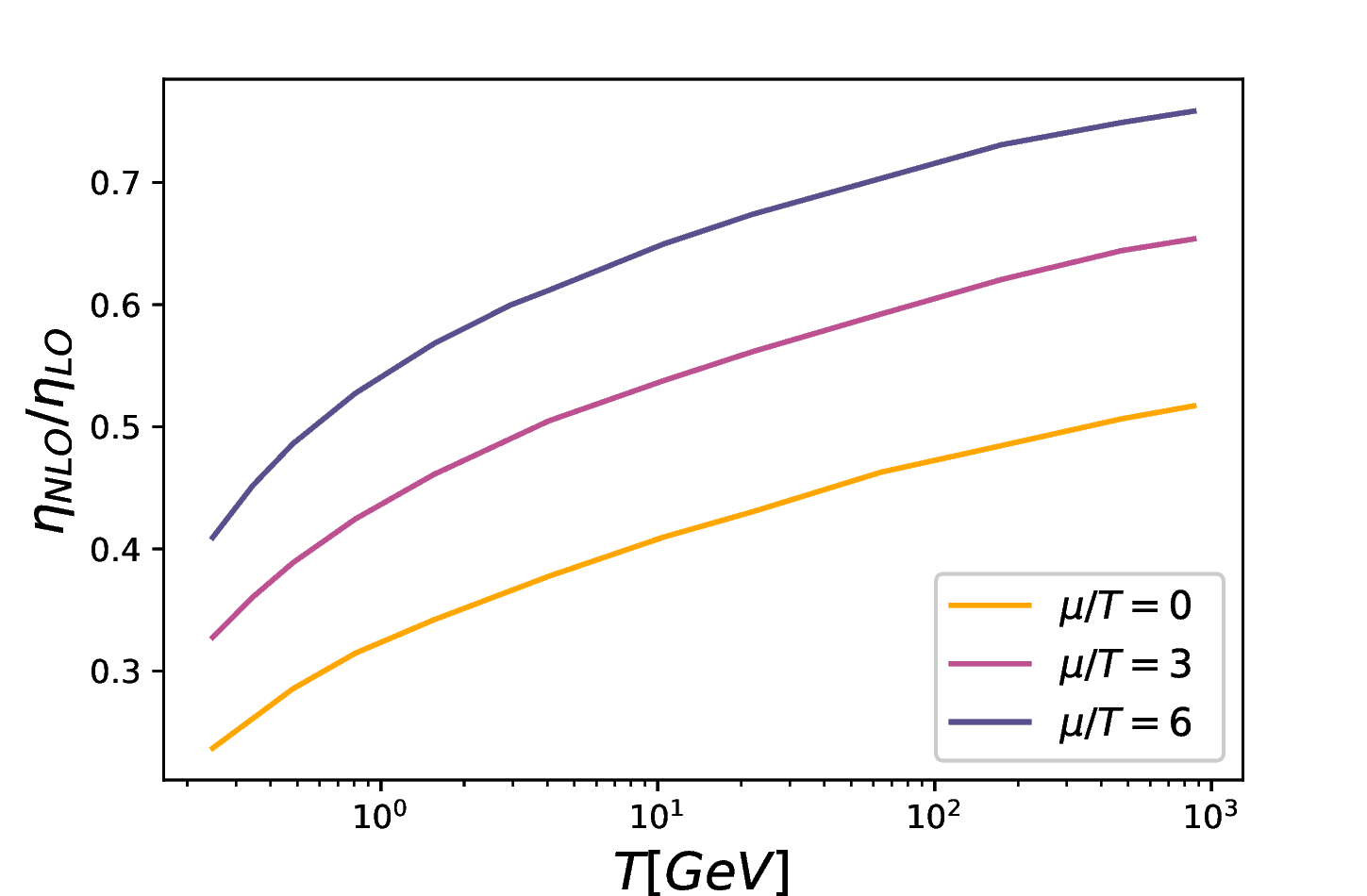}
    \caption{Ratio of the NLO shear viscosity to the LO shear viscosity, indicating the importance of subleading effects, for three chemical potentials over a range of temperatures.
    The renormalization scale was set at $\muEQCD=2.7T$ in all cases.}
    \label{visc_ratio}
\end{figure}

Since one motivation to study high-density regimes comes from the expectation that perturbation theory should show better convergence than at $\mu=0$, we present a more detailed comparison between LO and NLO results in Fig.~\ref{visc_ratio}.
We plot the ratio between LO and NLO calculations as a function of the temperature for each fixed chemical potential.
We reproduce the previous result \cite{Ghiglieri:2018dib} that the NLO corrections at $\mu=0$ are large, more than a factor of 2, up to $10^3$ GeV temperatures.
Once one goes to higher densities, the NLO effects become less dominant, falling below a factor of 2 by $T=500$ MeV for $\mu/T = 6$.
This suggests that the expansion in $g$ is better behaved at large $\mu/T$.
Nevertheless, NLO effects are rather important at all temperatures and chemical potentials we consider.



\section{Summary and Discussion}
\label{sec:summary}

We have extended the investigation of shear viscosity for hot QCD with chemical potentials to both leading and (almost) next-to-leading order.
We find that the viscosity-to-entropy ratio $\eta/s$ is larger at finite $\mu$ than at zero $\mu$ for all temperatures we consider.
However, the kinematic viscosity $\eta T/(e+P)$ is more weakly dependent on $\mu$, falling below the $\mu=0$ result except at NLO below $T=1$ GeV, where the $\mu/T$ dependence almost vanishes.
We argued previously that, at a given temperature or coupling strength, the convergence of perturbation theory should be better in the large $\mu/T$ regime than for $\mu=0$.
Our results show that this is indeed the case, but the improvement is not dramatic.
NLO effects are a factor of 2 at $\mu/T=0$ for
temperatures of hundreds of GeV,
whereas for $\mu=6T$ they are a factor of 2 at about 500 MeV temperature.
Perturbation theory does not converge at experimentally achievable combinations of $(\mu,T)$, but it comes closer at high $\mu$ than at small $\mu$.

It would be straightforward to apply the techniques developed here to the more physical case in which different quark species experience different chemical potentials.
This could allow the development of an interpolating estimate for the viscosity based on experimental data, hadron-resonance modeling, and NLO perturbative input, similar to what was recently developed by one of us \cite{Danhoni:2024kgi}.

\section*{Acknowledgements}
The authors acknowledge the support by the State of Hesse within the Research Cluster ELEMENTS (Project ID 500/10.006), and support by the Deutsche Forschungsgemeinschaft (DFG, German Research Foundation) through the CRC-TR 211 'Strong-interaction matter under extreme conditions'– project number 315477589 – TRR 211. The authors also thank Nicolas Wink, Jacopo Ghiglieri, and Kevin Ingles for instructive conversations.

\appendix

\section{Self-energies beyond the hard thermal loop approximation}
\label{AppA}
As discussed in \cite{Arnold:2002zm}, the t-channel exchange process is subtle, since the use of vacuum matrix elements leads to logarithmic divergences in the ``transport'' cross-section and in the inverse viscosity.
Baym \textsl{et al} already identified the correct procedure to fix this problem in the context of viscosity \cite{Baym:1990uj}; one should amend the matrix element through the inclusion of the thermal self-energy.
At the leading-log level, relevant in Refs.~\cite{Baym:1990uj,Arnold:2000dr}, it is sufficient to identify the energy scale where this occurs.
At leading order it is necessary to really include the self-energy, at least within an approximation which is adequate in the formal kinematic regime $(p,k) \sim T$ but $(q,\omega) \sim gT \ll T$.
This is precisely the regime where the HTL approximation should be valid.

At large chemical potentials we can try to do better.
In this regime, the self-energy is dominated by fermionic loops.
For these, the self-energy is gauge invariant and the full expression, not just the hard-loop limit, is known \cite{Klimov1,Weldon1}.
Specifically, the medium-induced part of the fermion-loop contribution to the gluonic self-energy is
\cite{Klimov1,Weldon1}
\begin{equation}
\label{PiGH}
\PiL = \frac{g^2 \nf T^2}{3}H(\omega,q), \hspace{0.5cm}
\PiT = \frac{g^2 \nf T^2}{3}\left(-\frac{1}{2}H(\omega,q) + \frac{q^2}{2(q^2-\omega^2)}G(\omega,q)\right)
\end{equation}
where $H(\omega,q)$ and $G(\omega,q)$ are, writing $2\omega_+=\omega+q$ and $2\omega_-=\omega-q$:
\begin{align}
\label{GH}
    \nonumber
    \Re G(\omega,q)&= \frac{3}{\pi^2T^2}\int_0^\infty dk f(k)\left(4k+\frac{q^2-\omega^2}{2q}\ln{\left[\frac{(k+\omega_-)(k-\omega_+)}{(k-\omega_-)(k+\omega_+)}\right]}\right)\,\\
    \nonumber
    \Re H(\omega,q)&= \frac{3}{\pi^2T^2}\int_0^\infty dk f(k)\Big(2k-\frac{4k^2+\omega^2-q^2}{4q}\ln{\left[\frac{(k+\omega_-)(k-\omega_+)}{(k-\omega_-)(k+\omega_+)}\right]}\,\\
    \nonumber
    &-\frac{2k\omega}{q}\ln{\frac{\omega_+}{\omega_-}}-\frac{k\omega}{q}\ln{\left[\frac{(k+\omega_-)(k-\omega_+)}{(k-\omega_-)(k+\omega_+)}\right]}\Big)\,\\
    \nonumber
    \Im G(\omega,q)&= \frac{3}{\pi^2T^2}\int_0^\infty dk f(k)\frac{q^2-\omega^2}{2q}[\Theta(k+\omega_+)\Theta(-k-\omega_-)-\Theta(-k+\omega_+)\Theta(k-\omega_-)]\,\\
    \nonumber
    \Im H(\omega,q)&= \frac{3}{\pi^2T^2}\int_0^\infty dk f(k)\Big[-\frac{(2k+\omega)^2-q^2}{4q}\Theta(k+\omega_+)\Theta(-k-\omega_-)\\
    &+\frac{(2k-\omega)^2-q^2}{4q}\Theta(k-\omega_-)\Theta(-k+\omega_+)+\frac{2k\omega}{q}\Theta(\omega_+)\Theta(-\omega_-)\Big]
\end{align}
The references only present the real parts, but the imaginary parts are easily obtained by using the retarded prescription $\omega \to \omega+i0$ and carefully interpreting the logarithms.
The hard thermal loop approximation corresponds to expanding the $k$ integrals in the approximation $k \gg (q,\omega)$, in which case the functions simplify:
\begin{align}
\label{GH-HTL}
    &H_{HTL}(\omega/q) = 1 - \frac{\omega}{2q}\ln{\frac{\omega_+}{\omega_-}}+i\frac{\pi\omega}{2q}\Theta(\omega_+)\Theta(-\omega_-)\\
    &G_{HTL}(\omega/q) = 1 .
\end{align}

One concern regarding the AMY leading-order calculation \cite{Arnold:2003zc} has always been the use of hard thermal loop self-energies, rather than full thermal self-energies.
At high chemical potential, where the hard loop is dominated by fermions, the above self-energies accurately represent the correct loop screening effect at any energy and momentum, without the need for an HTL approximation.
Therefore we can find out how well the HTL approximation works by comparing how the leading-order shear viscosity changes when we ``improve'' the screening effects by replacing the HTL approximation, \Eq{GH-HTL}, with the full self-energy, \Eq{GH}.

\begin{figure}[htb]
    \centering
    \includegraphics[scale=0.75]{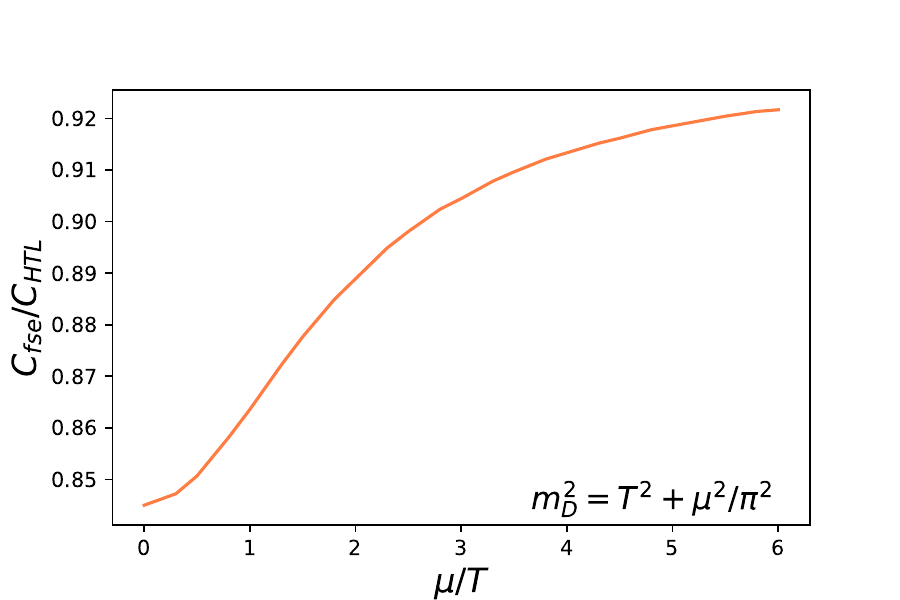}
    \caption{Ratio of diagram (C), integrated over the 1-parameter Ansatz, for the full vs the HTL self-energy.  The Debye mass is varied such that the coupling $g^2$ is kept fixed.}
    \label{htl}
\end{figure}

We make this comparison in figure \ref{htl}.
The figure considers a single collision process, diagram (C), and computes the collision integral $(\chi_{ij},\cc \chi_{ij})$ for the single-parameter Ansatz.
We then take the ratio of this collision rate for the case of the full self-energy over the collision rate using the HTL self-energy.
The Debye screening strength is chosen to vary with $\mu$ as it would if the coupling strength $g^2$ is held fixed.
The figure shows, first and foremost, that the effect of replacing the HTL self-energy with the full self-energy is not that large.
Also, the HTL approximation actually becomes more reliable as the chemical potential is increased.
This indicates that the use of the HTL approximation for the self-energies is justified and represents a smaller effect than the NLO (and, presumably, unknown NNLO) corrections.

\bibliographystyle{JHEP2}
\bibliography{refs}

\providecommand{\href}[2]{#2}\begingroup\raggedright\begin{thebibliography}{10}

\bibitem{Grefa:2022sav}
J.~Grefa, M.~Hippert, J.~Noronha, J.~Noronha-Hostler, I.~Portillo, C.~Ratti et~al., \emph{{Transport coefficients of the quark-gluon plasma at the critical point and across the first-order line}}, \href{https://doi.org/10.1103/PhysRevD.106.034024}{\emph{Phys. Rev. D} {\bfseries 106} (2022) 034024} [\href{https://arxiv.org/abs/2203.00139}{{\ttfamily 2203.00139}}].

\bibitem{PHENIX:2004vcz}
{\scshape PHENIX} collaboration, \emph{{Formation of dense partonic matter in relativistic nucleus-nucleus collisions at RHIC: Experimental evaluation by the PHENIX collaboration}}, \href{https://doi.org/10.1016/j.nuclphysa.2005.03.086}{\emph{Nucl. Phys. A} {\bfseries 757} (2005) 184} [\href{https://arxiv.org/abs/nucl-ex/0410003}{{\ttfamily nucl-ex/0410003}}].

\bibitem{STAR:2005gfr}
{\scshape STAR} collaboration, \emph{{Experimental and theoretical challenges in the search for the quark gluon plasma: The STAR Collaboration's critical assessment of the evidence from RHIC collisions}}, \href{https://doi.org/10.1016/j.nuclphysa.2005.03.085}{\emph{Nucl. Phys. A} {\bfseries 757} (2005) 102} [\href{https://arxiv.org/abs/nucl-ex/0501009}{{\ttfamily nucl-ex/0501009}}].

\bibitem{ALICE:2010suc}
{\scshape ALICE} collaboration, \emph{{Elliptic flow of charged particles in Pb-Pb collisions at 2.76 TeV}}, \href{https://doi.org/10.1103/PhysRevLett.105.252302}{\emph{Phys. Rev. Lett.} {\bfseries 105} (2010) 252302} [\href{https://arxiv.org/abs/1011.3914}{{\ttfamily 1011.3914}}].

\bibitem{Heinz:2013th}
U.~Heinz and R.~Snellings, \emph{{Collective flow and viscosity in relativistic heavy-ion collisions}}, \href{https://doi.org/10.1146/annurev-nucl-102212-170540}{\emph{Ann. Rev. Nucl. Part. Sci.} {\bfseries 63} (2013) 123} [\href{https://arxiv.org/abs/1301.2826}{{\ttfamily 1301.2826}}].

\bibitem{Luzum:2013yya}
M.~Luzum and H.~Petersen, \emph{{Initial State Fluctuations and Final State Correlations in Relativistic Heavy-Ion Collisions}}, \href{https://doi.org/10.1088/0954-3899/41/6/063102}{\emph{J. Phys. G} {\bfseries 41} (2014) 063102} [\href{https://arxiv.org/abs/1312.5503}{{\ttfamily 1312.5503}}].

\bibitem{JETSCAPE:2020mzn}
{\scshape JETSCAPE} collaboration, \emph{{Multisystem Bayesian constraints on the transport coefficients of QCD matter}}, \href{https://doi.org/10.1103/PhysRevC.103.054904}{\emph{Phys. Rev. C} {\bfseries 103} (2021) 054904} [\href{https://arxiv.org/abs/2011.01430}{{\ttfamily 2011.01430}}].

\bibitem{Laine:2003ay}
M.~Laine, \emph{{What is the simplest effective approach to hot QCD thermodynamics?}},  in \emph{{5th Internationa Conference on Strong and Electroweak Matter}}, pp.~137--146, 2003, \href{https://arxiv.org/abs/hep-ph/0301011}{{\ttfamily hep-ph/0301011}}, \href{https://doi.org/10.1142/9789812704498_0013}{DOI}.

\bibitem{Hietanen:2008tv}
A.~Hietanen, K.~Kajantie, M.~Laine, K.~Rummukainen and Y.~Schroder, \emph{{Three-dimensional physics and the pressure of hot QCD}}, \href{https://doi.org/10.1103/PhysRevD.79.045018}{\emph{Phys. Rev. D} {\bfseries 79} (2009) 045018} [\href{https://arxiv.org/abs/0811.4664}{{\ttfamily 0811.4664}}].

\bibitem{Goy:2016egl}
V.~A. Goy, V.~Bornyakov, D.~Boyda, A.~Molochkov, A.~Nakamura, A.~Nikolaev et~al., \emph{{Sign problem in finite density lattice QCD}}, \href{https://doi.org/10.1093/ptep/ptx018}{\emph{PTEP} {\bfseries 2017} (2017) 031D01} [\href{https://arxiv.org/abs/1611.08093}{{\ttfamily 1611.08093}}].

\bibitem{STAR:2014egu}
{\scshape STAR} collaboration, \emph{{Beam energy dependence of moments of the net-charge multiplicity distributions in Au+Au collisions at RHIC}}, \href{https://doi.org/10.1103/PhysRevLett.113.092301}{\emph{Phys. Rev. Lett.} {\bfseries 113} (2014) 092301} [\href{https://arxiv.org/abs/1402.1558}{{\ttfamily 1402.1558}}].

\bibitem{HADES:2019auv}
{\scshape HADES} collaboration, \emph{{Probing dense baryon-rich matter with virtual photons}}, \href{https://doi.org/10.1038/s41567-019-0583-8}{\emph{Nature Phys.} {\bfseries 15} (2019) 1040}.

\bibitem{Friese:2006dj}
V.~Friese, \emph{{The CBM experiment at GSI/FAIR}}, \href{https://doi.org/10.1016/j.nuclphysa.2006.06.018}{\emph{Nucl. Phys. A} {\bfseries 774} (2006) 377}.

\bibitem{Danhoni:2022xmt}
I.~Danhoni and G.~D. Moore, \emph{{Hot and dense QCD shear viscosity at leading log}}, \href{https://doi.org/10.1007/JHEP02(2023)124}{\emph{JHEP} {\bfseries 02} (2023) 124} [\href{https://arxiv.org/abs/2212.02325}{{\ttfamily 2212.02325}}].

\bibitem{Ghiglieri:2018dib}
J.~Ghiglieri, G.~D. Moore and D.~Teaney, \emph{{QCD Shear Viscosity at (almost) NLO}}, \href{https://doi.org/10.1007/JHEP03(2018)179}{\emph{JHEP} {\bfseries 03} (2018) 179} [\href{https://arxiv.org/abs/1802.09535}{{\ttfamily 1802.09535}}].

\bibitem{Danhoni:2024kgi}
I.~Danhoni, J.~S.~S. Martin and J.~Noronha-Hostler, \emph{{Shear viscosity from perturbative Quantum Chromodynamics to the hadron resonance gas at finite baryon, strangeness, and electric charge densities}},  \href{https://arxiv.org/abs/2406.04968}{{\ttfamily 2406.04968}}.

\bibitem{Arnold:2003zc}
P.~B. Arnold, G.~D. Moore and L.~G. Yaffe, \emph{{Transport coefficients in high temperature gauge theories. 2. Beyond leading log}}, \href{https://doi.org/10.1088/1126-6708/2003/05/051}{\emph{JHEP} {\bfseries 05} (2003) 051} [\href{https://arxiv.org/abs/hep-ph/0302165}{{\ttfamily hep-ph/0302165}}].

\bibitem{Arnold:2000dr}
P.~B. Arnold, G.~D. Moore and L.~G. Yaffe, \emph{{Transport coefficients in high temperature gauge theories. 1. Leading log results}}, \href{https://doi.org/10.1088/1126-6708/2000/11/001}{\emph{JHEP} {\bfseries 11} (2000) 001} [\href{https://arxiv.org/abs/hep-ph/0010177}{{\ttfamily hep-ph/0010177}}].

\bibitem{Landau1987Fluid}
L.~D. Landau and E.~M. Lifshitz, \emph{Fluid Mechanics, Second Edition: Volume 6 (Course of Theoretical Physics)}, Course of theoretical physics / by L. D. Landau and E. M. Lifshitz, Vol. 6. Butterworth-Heinemann, 1987.

\bibitem{Arnold:2006fz}
P.~B. Arnold, C.~Dogan and G.~D. Moore, \emph{{The Bulk Viscosity of High-Temperature QCD}}, \href{https://doi.org/10.1103/PhysRevD.74.085021}{\emph{Phys. Rev. D} {\bfseries 74} (2006) 085021} [\href{https://arxiv.org/abs/hep-ph/0608012}{{\ttfamily hep-ph/0608012}}].

\bibitem{Jeon:1994if}
S.~Jeon, \emph{{Hydrodynamic transport coefficients in relativistic scalar field theory}}, \href{https://doi.org/10.1103/PhysRevD.52.3591}{\emph{Phys. Rev. D} {\bfseries 52} (1995) 3591} [\href{https://arxiv.org/abs/hep-ph/9409250}{{\ttfamily hep-ph/9409250}}].

\bibitem{Gagnon:2006hi}
J.-S. Gagnon and S.~Jeon, \emph{{Leading order calculation of electric conductivity in hot quantum electrodynamics from diagrammatic methods}}, \href{https://doi.org/10.1103/PhysRevD.75.025014}{\emph{Phys. Rev. D} {\bfseries 75} (2007) 025014} [\href{https://arxiv.org/abs/hep-ph/0610235}{{\ttfamily hep-ph/0610235}}].

\bibitem{Gagnon:2007qt}
J.-S. Gagnon and S.~Jeon, \emph{{Leading Order Calculation of Shear Viscosity in Hot Quantum Electrodynamics from Diagrammatic Methods}}, \href{https://doi.org/10.1103/PhysRevD.76.105019}{\emph{Phys. Rev. D} {\bfseries 76} (2007) 105019} [\href{https://arxiv.org/abs/0708.1631}{{\ttfamily 0708.1631}}].

\bibitem{Arnold:2002zm}
P.~B. Arnold, G.~D. Moore and L.~G. Yaffe, \emph{{Effective kinetic theory for high temperature gauge theories}}, \href{https://doi.org/10.1088/1126-6708/2003/01/030}{\emph{JHEP} {\bfseries 01} (2003) 030} [\href{https://arxiv.org/abs/hep-ph/0209353}{{\ttfamily hep-ph/0209353}}].

\bibitem{Baym:1990uj}
G.~Baym, H.~Monien, C.~J. Pethick and D.~G. Ravenhall, \emph{{Transverse Interactions and Transport in Relativistic Quark - Gluon and Electromagnetic Plasmas}}, \href{https://doi.org/10.1103/PhysRevLett.64.1867}{\emph{Phys. Rev. Lett.} {\bfseries 64} (1990) 1867}.

\bibitem{Jeon:1995zm}
S.~Jeon and L.~G. Yaffe, \emph{{From quantum field theory to hydrodynamics: Transport coefficients and effective kinetic theory}}, \href{https://doi.org/10.1103/PhysRevD.53.5799}{\emph{Phys. Rev. D} {\bfseries 53} (1996) 5799} [\href{https://arxiv.org/abs/hep-ph/9512263}{{\ttfamily hep-ph/9512263}}].

\bibitem{Braaten:1989mz}
E.~Braaten and R.~D. Pisarski, \emph{{Soft Amplitudes in Hot Gauge Theories: A General Analysis}}, \href{https://doi.org/10.1016/0550-3213(90)90508-B}{\emph{Nucl. Phys. B} {\bfseries 337} (1990) 569}.

\bibitem{Frenkel:1989br}
J.~Frenkel and J.~C. Taylor, \emph{{High Temperature Limit of Thermal QCD}}, \href{https://doi.org/10.1016/0550-3213(90)90661-V}{\emph{Nucl. Phys. B} {\bfseries 334} (1990) 199}.

\bibitem{Kurkela:2018oqw}
A.~Kurkela and A.~Mazeliauskas, \emph{{Chemical equilibration in weakly coupled QCD}}, \href{https://doi.org/10.1103/PhysRevD.99.054018}{\emph{Phys. Rev. D} {\bfseries 99} (2019) 054018} [\href{https://arxiv.org/abs/1811.03068}{{\ttfamily 1811.03068}}].

\bibitem{Moore:2001fga}
G.~D. Moore, \emph{{Transport coefficients in large N(f) gauge theory: Testing hard thermal loops}}, \href{https://doi.org/10.1088/1126-6708/2001/05/039}{\emph{JHEP} {\bfseries 05} (2001) 039} [\href{https://arxiv.org/abs/hep-ph/0104121}{{\ttfamily hep-ph/0104121}}].

\bibitem{Klimov1}
V.~V. Klimov, \emph{{Collective Excitations in a Hot Quark Gluon Plasma}}, {\emph{Sov. Phys. JETP} {\bfseries 55} (1982) 199}.

\bibitem{Weldon1}
H.~A. Weldon, \emph{{Covariant Calculations at Finite Temperature: The Relativistic Plasma}}, \href{https://doi.org/10.1103/PhysRevD.26.1394}{\emph{Phys. Rev. D} {\bfseries 26} (1982) 1394}.

\bibitem{Ghiglieri:2015ala}
J.~Ghiglieri, G.~D. Moore and D.~Teaney, \emph{{Jet-Medium Interactions at NLO in a Weakly-Coupled Quark-Gluon Plasma}}, \href{https://doi.org/10.1007/JHEP03(2016)095}{\emph{JHEP} {\bfseries 03} (2016) 095} [\href{https://arxiv.org/abs/1509.07773}{{\ttfamily 1509.07773}}].

\bibitem{Ghiglieri:2015zma}
J.~Ghiglieri and D.~Teaney, \emph{{Parton energy loss and momentum broadening at NLO in high temperature QCD plasmas}}, \href{https://doi.org/10.1142/S0218301315300131}{\emph{Int. J. Mod. Phys. E} {\bfseries 24} (2015) 1530013} [\href{https://arxiv.org/abs/1502.03730}{{\ttfamily 1502.03730}}].

\bibitem{Moore:2004tg}
G.~D. Moore and D.~Teaney, \emph{{How much do heavy quarks thermalize in a heavy ion collision?}}, \href{https://doi.org/10.1103/PhysRevC.71.064904}{\emph{Phys. Rev. C} {\bfseries 71} (2005) 064904} [\href{https://arxiv.org/abs/hep-ph/0412346}{{\ttfamily hep-ph/0412346}}].

\bibitem{CaronHuot:2008ni}
S.~Caron-Huot, \emph{{O(g) plasma effects in jet quenching}}, \href{https://doi.org/10.1103/PhysRevD.79.065039}{\emph{Phys. Rev. D} {\bfseries 79} (2009) 065039} [\href{https://arxiv.org/abs/0811.1603}{{\ttfamily 0811.1603}}].

\bibitem{Aurenche:2002pd}
P.~Aurenche, F.~Gelis and H.~Zaraket, \emph{{A Simple sum rule for the thermal gluon spectral function and applications}}, \href{https://doi.org/10.1088/1126-6708/2002/05/043}{\emph{JHEP} {\bfseries 05} (2002) 043} [\href{https://arxiv.org/abs/hep-ph/0204146}{{\ttfamily hep-ph/0204146}}].

\bibitem{Ghiglieri:2013gia}
J.~Ghiglieri, J.~Hong, A.~Kurkela, E.~Lu, G.~D. Moore and D.~Teaney, \emph{{Next-to-leading order thermal photon production in a weakly coupled quark-gluon plasma}}, \href{https://doi.org/10.1007/JHEP05(2013)010}{\emph{JHEP} {\bfseries 05} (2013) 010} [\href{https://arxiv.org/abs/1302.5970}{{\ttfamily 1302.5970}}].

\bibitem{Ghisoiu:2014mha}
I.~Ghisoiu and M.~Laine, \emph{{Interpolation of hard and soft dilepton rates}}, \href{https://doi.org/10.1007/JHEP10(2014)083}{\emph{JHEP} {\bfseries 10} (2014) 083} [\href{https://arxiv.org/abs/1407.7955}{{\ttfamily 1407.7955}}].

\bibitem{Ghiglieri:2014kma}
J.~Ghiglieri and G.~D. Moore, \emph{{Low Mass Thermal Dilepton Production at NLO in a Weakly Coupled Quark-Gluon Plasma}}, \href{https://doi.org/10.1007/JHEP12(2014)029}{\emph{JHEP} {\bfseries 12} (2014) 029} [\href{https://arxiv.org/abs/1410.4203}{{\ttfamily 1410.4203}}].

\bibitem{Braaten:1991gm}
E.~Braaten and R.~D. Pisarski, \emph{{Simple effective Lagrangian for hard thermal loops}}, \href{https://doi.org/10.1103/PhysRevD.45.R1827}{\emph{Phys. Rev. D} {\bfseries 45} (1992) R1827}.

\end{thebibliography}\endgroup
\end{document}